\documentclass[twoside,twocolumn,9pt]{article}
\usepackage{extsizes}
\usepackage[super,sort&compress,comma]{natbib}
\usepackage[version=3]{mhchem}
\usepackage[left=1.5cm, right=1.5cm, top=1.785cm, bottom=2.0cm]{geometry}
\usepackage{balance}
\usepackage{mathptmx}
\usepackage{sectsty}
\usepackage{graphicx}
\usepackage{lastpage}
\usepackage[format=plain,justification=justified,singlelinecheck=false,font={stretch=1.125,small,sf},labelfont=bf,labelsep=space]{caption}
\usepackage{float}
\usepackage{fancyhdr}
\usepackage{fnpos}
\usepackage[english]{babel}
\addto{\captionsenglish}{%
  \renewcommand{\refname}{Notes and references}
}
\usepackage{array}
\usepackage{droidsans}
\usepackage{charter}
\usepackage[T1]{fontenc}
\usepackage[usenames,dvipsnames]{xcolor}
\usepackage{setspace}
\usepackage[compact]{titlesec}
\usepackage{hyperref}

\usepackage{epstopdf}

\definecolor{cream}{RGB}{222,217,201}
\usepackage{amsmath,amssymb,amsthm,url}
\usepackage{comment}
\DeclareMathOperator{\Dom}{Dom}

\usepackage{xcolor}

\begin{document}

\pagestyle{fancy}
\thispagestyle{plain}
\fancypagestyle{plain}{
\renewcommand{\headrulewidth}{0pt}
}

\makeFNbottom
\makeatletter
\renewcommand\LARGE{\@setfontsize\LARGE{15pt}{17}}
\renewcommand\Large{\@setfontsize\Large{12pt}{14}}
\renewcommand\large{\@setfontsize\large{10pt}{12}}
\renewcommand\footnotesize{\@setfontsize\footnotesize{7pt}{10}}
\makeatother

\renewcommand{\thefootnote}{\fnsymbol{footnote}}
\renewcommand\footnoterule{\vspace*{1pt}%
\color{cream}\hrule width 3.5in height 0.4pt \color{black}\vspace*{5pt}}
\setcounter{secnumdepth}{5}

\makeatletter
\renewcommand\@biblabel[1]{#1}
\renewcommand\@makefntext[1]%
{\noindent\makebox[0pt][r]{\@thefnmark\,}#1}
\makeatother
\renewcommand{\figurename}{\small{Fig.}~}
\sectionfont{\sffamily\Large}
\subsectionfont{\normalsize}
\subsubsectionfont{\bf}
\setstretch{1.125} 
\setlength{\skip\footins}{0.8cm}
\setlength{\footnotesep}{0.25cm}
\setlength{\jot}{10pt}
\titlespacing*{\section}{0pt}{4pt}{4pt}
\titlespacing*{\subsection}{0pt}{15pt}{1pt}

\fancyfoot{}
\fancyfoot[LO,RE]{\vspace{-7.1pt}\includegraphics[height=9pt]{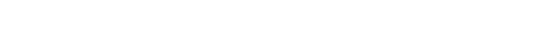}}
\fancyfoot[CO]{\vspace{-7.1pt}\hspace{13.2cm}\includegraphics{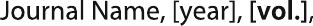}}
\fancyfoot[CE]{\vspace{-7.2pt}\hspace{-14.2cm}\includegraphics{head_foot/RF}}
\fancyfoot[RO]{\footnotesize{\sffamily{1--\pageref{LastPage} ~\textbar  \hspace{2pt}\thepage}}}
\fancyfoot[LE]{\footnotesize{\sffamily{\thepage~\textbar\hspace{3.45cm} 1--\pageref{LastPage}}}}
\fancyhead{}
\renewcommand{\headrulewidth}{0pt}
\renewcommand{\footrulewidth}{0pt}
\setlength{\arrayrulewidth}{1pt}
\setlength{\columnsep}{6.5mm}
\setlength\bibsep{1pt}

\makeatletter
\newlength{\figrulesep}
\setlength{\figrulesep}{0.5\textfloatsep}

\newcommand{\topfigrule}{\vspace*{-1pt}%
\noindent{\color{cream}\rule[-\figrulesep]{\columnwidth}{1.5pt}} }

\newcommand{\botfigrule}{\vspace*{-2pt}%
\noindent{\color{cream}\rule[\figrulesep]{\columnwidth}{1.5pt}} }

\newcommand{\dblfigrule}{\vspace*{-1pt}%
\noindent{\color{cream}\rule[-\figrulesep]{\textwidth}{1.5pt}} }

\makeatother
\newcommand*{\emp}[1]{\textbf{#1}}
\newcommand*{\mma}[1]{\textbf{\texttt{#1}}}

\newcommand*{\nulb}{\boldsymbol{0}}
\newcommand*{\oneb}{\boldsymbol{1}}
\newcommand*{\alphab}{\boldsymbol{\alpha}}
\newcommand*{\betab}{\boldsymbol{\beta}}
\newcommand*{\gammab}{\boldsymbol{\gamma}}
\newcommand*{\mub}{\boldsymbol{\mu}}
\newcommand*{\xib}{\boldsymbol{\xi}}

\newcommand*{\Ab}{\mathbf{A}}
\newcommand*{\Nb}{\mathbf{N}}
\newcommand*{\Wb}{\mathbf{W}}
\newcommand*{\cb}{\mathbf{c}}
\newcommand*{\fb}{\mathbf{f}}
\newcommand*{\kb}{\mathbf{k}}
\newcommand*{\nb}{\mathbf{n}}
\newcommand*{\xb}{\mathbf{x}}

\newcommand*{\N}{\mathbb{N}}
\newcommand*{\R}{\mathbb{R}}

\newcommand*{\CC}{\mathcal{C}}
\newcommand*{\MM}{\mathcal{M}}
\newcommand*{\RR}{\mathcal{R}}

\newcommand*{\ikde}{induced kinetic differential equation}
\newcommand*{\Lit}{\mathrm{dm}^3}
\newcommand*{\Litsix}{\mathrm{dm}^6}
\newcommand*{\mL}{\mol\;\mathrm{dm}^{-3}}
\newcommand*{\mol}{\,\mathrm{mol}}
\newcommand*{\rateb}{\mathbf{rate}}
\newcommand*{\mth}{m^{\mathrm{th}}}
\newcommand*{\rth}{r^{\mathrm{th}}}
\newcommand*{\T}{^{\top}}

\theoremstyle{definition}
\newtheorem{definition}{Definition}
\newtheorem{example}{Example}
\newtheorem{theorem}{Theorem}
\newtheorem{conjecture}{Conjecture}
\newtheorem{corollary}{Corollary}
\newtheorem{proposition}{Proposition}
\newcommand{\Pf}[1]{{\textit{Proof}\ }#1}

\newcommand{\vilmos}{\textcolor{blue}}
\newcommand{\torol}{\textcolor{red}}

\twocolumn[
  \begin{@twocolumnfalse}
{\includegraphics[height=30pt]{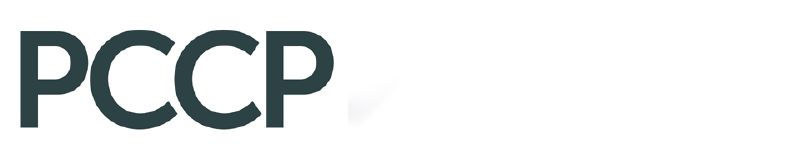}\hfill\raisebox{0pt}[0pt][0pt]{\includegraphics[height=55pt]{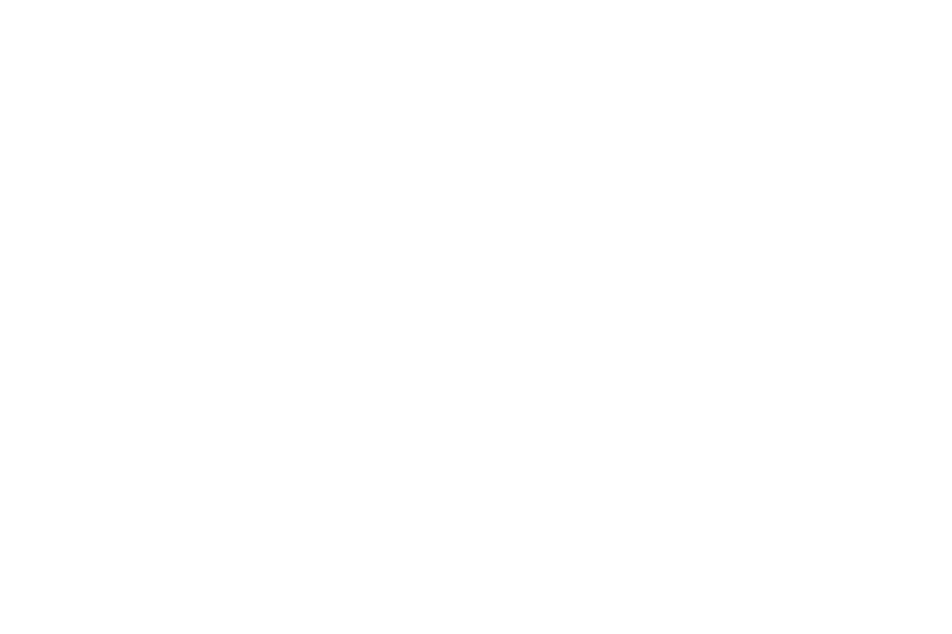}}\\[1ex]
\includegraphics[width=18.5cm]{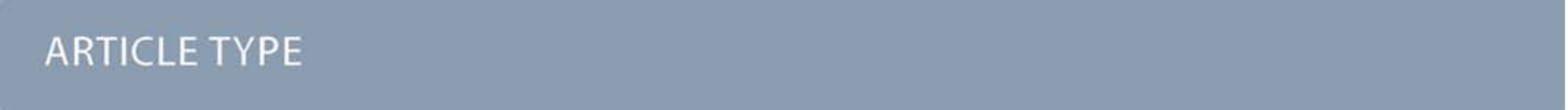}}\par
\vspace{1em}
\sffamily
\begin{tabular}{m{4.5cm} p{13.5cm} }

\includegraphics{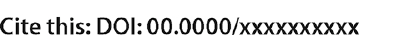} & \noindent\LARGE{\textbf{Reaction extent or advancement of reaction: A definition for complex chemical reactions$^{\dag,\ddag}$}}\\ 
\vspace{0.3cm} & \vspace{0.3cm} \\

 & \noindent\large{Vilmos G\'asp\'ar$^{\ast}$\textit{$^{a}$} and J\'anos T\'oth\textit{$^{b,c}$}} \\

\includegraphics{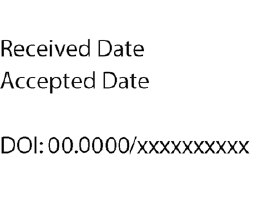} & \noindent\normalsize{The concept of reaction extent (the progress of a reaction, advancement of the reaction, conversion, etc.) was introduced around 100 years ago. Most of the literature provides a definition for the exceptional case of a single reaction step or gives an implicit definition that cannot be made explicit. 
There are views that the reaction extent somehow has to tend to 1 when the reaction goes to completion as time tends to infinity. 
However, there is no agreement on which function should tend to 1. 
Starting from the standard definition by IUPAC and following the classical works by De Donder, Aris, and Croce we extend the classic definition of the reaction extent for an arbitrary number of species and reaction steps. 
The new general, explicit definition is also valid for non-mass action kinetics. 
We also studied the mathematical properties (evolution equation, continuity, monotony, differentiability, etc.) of the defined quantity, connecting them to the formalism of modern reaction kinetics. 
Our approach tries to adhere to the customs of chemists and be mathematically correct simultaneously. To make the exposition easy to understand, we use simple chemical examples and many figures, throughout. We also show how to apply this concept to exotic reactions: reactions with more than one stationary state, oscillatory reactions, and reactions showing chaotic behavior.
The main advantage of the new definition of reaction extent is that by knowing the kinetic model of a reacting system one can now calculate not only the time evolution of the concentration of each reacting species but also the number of occurrences of the individual reaction events.} 

\\

\end{tabular}

 \end{@twocolumnfalse} \vspace{0.6cm}
  
  ]

\renewcommand*\rmdefault{bch}\normalfont\upshape
\rmfamily
\section*{}
\vspace{-1cm}



\footnotetext{\textit{$^{a}$~Laboratory of Nonlinear Chemical Dynamics, Institute of Chemistry, ELTE E\"otv\"os Lor\'and University, Budapest, Hungary.}}
\footnotetext{\textit{$^{b}$~Budapest University of Technology and Economics, Department of Analysis, Budapest, Hungary. Fax: +361 463 3172; Tel: +361 463 2314; E-mail: jtoth@math.bme.hu}}
\footnotetext{\textit{$^{c}$~Chemical Kinetics Laboratory, Institute of Chemistry, ELTE E\"otv\"os Lor\'and University, Budapest, Hungary.}}

\footnotetext{\dag~Electronic Supplementary Information (ESI) available: [details of any supplementary information available should be included here]. See DOI: 10.1039/cXCP00000x/}

\footnotetext{\ddag~Based on the talk given at the 2nd International Conference on Reaction Kinetics, Mechanisms and Catalysis. 20--22 May 2021, Budapest, Hungary}

\section{Introduction} 
The concept of reaction extent (most often denoted by \(\xi\)) is more than 100 years old \cite{dedondervanrysselberghe}. 
Its importance is emphasized by the mere fact that it has been included in the IUPAC Green Book\cite{millscvitashomankallaykuchitsu} (see page 43). 
Two definitions are given that are equivalent in the treated very simple special case of a
single reaction step:
\begin{equation}\label{eq:cm}
\ce{$\sum_{m=1}^M\alpha_{m}$X($m$) = $\sum_{m=1}^M\beta_{m}$X($m$)}; 
\end{equation}
where \(M\) is the number of chemical species \ce{X(1)}, \ce{X(2)}, \dots, \ce{X($M$)};
and the integers \(\alpha_m\) and \(\beta_m\) are the corresponding stoichiometric coefficients of the reactant and product species, respectively.
The first definition is:
\begin{equation}\label{eq:iupac1def}
 n_{\ce{X($m$)}}=n^0_{\ce{X($m$)}}+ \nu_{m}\xi,  
\end{equation}
where \ce{n_{\ce{X($m$)}}} and $n^0_{\ce{X($m$)}}$ are the actual and initial quantities (number of moles) of the species \ce{X($m$)}, respectively. The symbol \(\nu_m\) is the generalized stoichiometric number. It is negative for a reactant and positive for a product species.
The second definition is
\begin{equation}\label{eq:iupac2def}
\Delta\xi=\frac{\Delta n_{\ce{X($m$)}}}{\nu_m}=
\frac{n_{\ce{X($m$)}}-n^0_{\ce{X($m$)}}}{\nu_m}.
\end{equation}
A slightly different version is given in Ref.\cite{goldloeningmcnaughtsehmi} and by the electronic version \url{https://goldbook.iupac.org/terms/view/E02283} called IUPAC Gold Book\cite{mcnaughtwilkinson}:
\begin{equation}\label{eq:iupac3def}
\mathrm{d}\xi=\frac{\mathrm{d} n_{\ce{X($m$)}}}{\nu_m}.
\end{equation}
The above-cited definitions have been summarized in the book by Stepanov et al.\cite{stepanoverlikinafilippov}. The authors also give a good introduction to the methods of linear algebra applied in Reaction Kinetics.

With an eye on the applicability of the concept in modern formal reaction kinetics (or, chemical reaction network theory) 
as exposed by Feinberg\cite{feinbergbook} and Tóth et al.\cite{tothnagypapp} the following points seem crucial:
\begin{enumerate}
\item
Starting from the original definition by De Donder and Van Rysselberghe\cite{dedondervanrysselberghe}, we extend the definition to an \emp{arbitrary} number of reaction steps.
\item
We do not restrict ourselves to reversible steps.
\item
We do not require linear independence of the reaction steps.
\item
We do not "order the steps to one side" which would result in hiding the difference between the steps like \ce{X -> Y} and \ce{X + Y -> 2Y}.
\item
We do not consider and take into account the atomic (or molecular) structure of the species.
\item 
We do not use differentials when introducing the concept
(cf. p. 61. of Ref.\cite{truesdell}).
\item
We shall give an explicit definition more similar to Eq. \eqref{eq:iupac2def} rather than to Eq. \eqref{eq:iupac1def}.
\item
We take into consideration the volume of the reacting mixture to be able to calculate the number of individual reaction events.
\end{enumerate}

The structure of our paper is as follows. 
Section \ref{sec:concept} introduces the concept for reaction networks of arbitrary complexity: for any number of reaction steps and species, mass action kinetics is not assumed. 
As it is a usual requirement that the reaction extent tends to 1 when "the reaction tends to its end", we try to find quantities derived from our reaction extent having this property in Section \ref{sec:what}.
It will turn out in many examples that the reaction extents do not tend to 1 in any sense. We show, however, that they contain quite relevant information about the time evolution of the reactions: they measure (or give) the number of the occurrences of the individual reaction events.
These examples will also reflect the fact that the reaction events do not cease  during equilibrium, and this can be seen without referring to fluctuations.
As the closure of our paper, we show applications of the concept
to more complicated cases: those with multiple stationary states, oscillation, and chaos.

In this part, first, we analyze the classical multi-stationary example
by Horn and Jackson\cite{hornjackson}. 
As to oscillatory reactions, we start with the irreversible Lotka--Volterra reaction, and we also study the reversible Lotka--Volterra reaction both in the detailed balanced and not detailed balanced cases. 
Our following oscillatory example will be an experimental system studied by Rábai\cite{rabai}. 
As a chaotic example, we shall take a slightly modified version of that oscillatory system. 
Discussion of the Conclusions and a list of Notations come last.
The proofs of the statements and Theorems are relegated to an Appendix so as to improve the logical flow of the manuscript without getting side-tracked.
Supporting Information is given in a PDF file; upon request, the corresponding Wolfram Language notebook---the source of the PDF file---will be provided to the interested reader.
\section{The concept of reaction extent}\label{sec:concept}
Starting from the classical works\cite{dedondervanrysselberghe,arisprolegomena1,croce}
and relying on the consensus of the chemists' community as formulated by Laidler\cite{laidlerglossary} our aim is to present a treatment more general than any of the definitions introduced and applied up to now.
\subsection{Motivation and fundamental definitions}
We are going to use the following concepts.
\subsubsection{Fundamental notations and definitions: The framework.}
Following the books by Feinberg\cite{feinbergbook} and by T\'oth et al.\cite{tothnagypapp} we consider a \emp{complex chemical reaction}, simply \emp{reaction}, or \emp{reaction network} as a set consisting of \emp{reaction steps} as follows: 
\begin{equation}\label{eq:ccr}
\left\{\ce{$\sum_{m=1}^M\alpha_{m,r}$X($m$) -> $\sum_{m=1}^M\beta_{m,r}$X($m$)}\quad (r=1,2,\dots,R)\right\}; 
\end{equation}
where 
\begin{enumerate}
\item 
the chemical \emp{species} are \ce{X($1$)}, \ce{X($2$)}, \dots, \ce{X($M$)}---take note that their quantities \(N_{\ce{X($m$)}}\) or \(N_m\) will be applied interchangeably;
\item
the \emp{reaction steps} are numbered from 1 to \(R;\) 
\item
here \(M\) and \(R\) are positive integers;
\item
\(\alphab:=[\alpha_{m,r}]\) and 
\(\betab:=[\beta_{m,r}]\) are \(M\times R\) matrices of non-negative integer components called \emp{stoichiometric coefficients}, with the properties that all the species take part in at least one reaction step (\(\forall m \exists r: \beta_{m,r}\neq\alpha_{m,r}\)), and all the reaction steps do have some effect (\(\forall r \exists m: \beta_{m,r}\neq\alpha_{m,r}\)), and finally
\item
\(\gammab:=\betab-\alphab\) is the \emp{stoichiometric matrix} of  \emp{stoichiometric numbers}.
\end{enumerate}
Instead of Eq. \eqref{eq:ccr} some authors prefer writing this:
\begin{equation}\label{eq:ccrreduced}
\ce{$\sum_{m=1}^M\gamma_{m,r}$X($m$) = 0}\quad (r=1,2,\dots,R).
\end{equation}
This formulation immediately excludes reaction steps like \ce{X + Y -> 2Y} (used e.g. to describe a step in the Lotka--Volterra reaction), or reduces it to \ce{X -> Y}, 
changing the stoichiometric coefficients used to formulate mass-action type kinetics. Similarly, an autocatalytic step that may be worth studying, see e.g. pp. 63 and 66 in the book by Aris\cite{arisintro}, like \ce{X -> 2X} appears oversimplified as \ce{0 -> X}. 
Another possibility is to exclude the \emp{empty complex}, implying involuntarily that we get rid of the possibility to simply represent in- and outflow with \ce{0 -> X} and \ce{X -> 0}, respectively. 
These last two examples evidently mean mass-creation and 
mass-destruction. 
If one does not like these one should explicitly say that one is only interested in mass-conserving reactions. 
Sometimes mass creation and mass destruction are slightly less obvious than above, 
see the reaction network
\begin{equation*}
\ce{X -> Y + U},\quad \ce{Z -> X + U},\quad \ce{Z -> U},
\end{equation*}
which is mass-producing.

It may happen that one would like to exclude reaction steps with more than two particles on the left side, such as 
\[
\ce{2MnO4- + 6H+ + 5H2C2O4 = 2Mn^2+ + 8H2O + 10CO2}. 
\]
Such steps do occur e.g. on page 1236 of Kov\'acs et al.\cite{kovacsvizvaririedeltoth} when dealing with \emp{overall reactions}. The theory and applications of decomposition of overall reactions into elementary steps\cite{kovacsvizvaririedeltoth,pappvizvari} would have been impossible without the framework of formal reaction kinetics. Someone may be interested in complex chemical reactions  consisting of reversible steps only. Then, they have to write down all the forward and the corresponding backward reaction steps.

Taking into consideration restrictions of the above kind usually 
does not make the mathematical treatment easier. 
Sometimes it needs hard work to figure out how they can be checked, 
as it is in the case of mass conservation of models containing species without atomic structure,\cite{deaktothvizvari,tothnagypapp} or in relation to the existence of oscillatory reactions.\cite{potalotka}
To sum up: an author has the right to make any restriction 
thought to be chemically important, but these restrictions should be declared at the outset. 
Finally, we mention our main assumption: all the steps in Eq. \eqref{eq:ccr} are really present, i.e. they proceed with a positive rate whenever the species on the left side are present.

We now provide a simple example to make the understanding easier.
\subsubsection{A simple example.}
Let us take an example that may be deemed chemically oversimplified but not too trivial, still simple enough so as not to be lost in the details. 
Assume that water formation follows the reversible reaction step 
\ce{2H2 + O2 <=> 2H2O}.
This means that we do not take into consideration either the atomic structure of the species, or the realistic details of water formation.
Let the forward step be represented in a more abstract way: \ce{2X + Y -> 2Z}.

The number of species, denoted as above by \(M\), is 3, and the number of (irreversible) reaction steps, denoted as above by \(R\), is 1. 
The \(3\times1\) stoichiometric matrix \(\gammab\) consisting of the  stoichiometric numbers is:
\(\begin{bmatrix}-2\\-1\\2\end{bmatrix}\). 
In case this step occurs five times, the vector of the 
\emp{number of individual species}
will change as follows:
\begin{equation*}
\begin{bmatrix}
N_{\ce{X}}-N_{\ce{X}}^0\\
N_{\ce{Y}}-N_{\ce{Y}}^0\\
N_{\ce{Z}}-N_{\ce{Z}}^0
\end{bmatrix}=
5\begin{bmatrix}
-2\\
-1\\
2
\end{bmatrix},
\end{equation*}
where \(N_{\ce{X}}^0\) is the number of molecules of species \ce{X} at the beginning, and \(N_{\ce{X}}\) is the number of molecules of species \ce{X} after five reaction events, and so on. 
If one considers the reversible reaction 
\ce{2X + Y <=> 2Z}, and assumes that the backward reaction step takes place three times then the total change is
\begin{equation}
\begin{bmatrix}
N_{\ce{X}}\\N_{\ce{Y}}\\N_{\ce{Z}}
\end{bmatrix}-
\begin{bmatrix}
N_{\ce{X}}^0\\N_{\ce{Y}}^0\\N_{\ce{Z}}^0
\end{bmatrix}=
5\begin{bmatrix}
-2\\-1\\2
\end{bmatrix}+
3\begin{bmatrix}
2\\1\\-2
\end{bmatrix}.\label{eq:spec}
\end{equation}
Note that both the number of molecules and the number of the occurrence of reaction events are positive integers. 

Eq. (2.2) of Ref.\cite{kurtz} is of the same form as our Eq. \eqref{eq:spec}. Kurtz is interested mainly in reversible and detailed balanced complex chemical reactions, and, more importantly, in the relationship of their deterministic and stochastic models. This is the reason why he formulates his Eq. (2.2) for the slightly restricted case only. As to the relationship between discrete and continuous descriptions, we follow here more or less Kurtz\cite{kurtz} and T\'oth et al.\cite{tothnagypapp} We cannot rely on a discrete state deterministic model of reaction kinetics---that would be desirable---because such a model does not exist as far as we know. 

\subsubsection{The general case.}
Before providing general definitions, we mention that Dumon et al.\cite{dumonlichanotpoquet} formulated a series of requirements that---according to them---should be obeyed by a well-defined reaction extent.
Unfortunately, we are unable to accept most of these requirements. 
Let us mention only one: the reaction extent should be independent of the choice of stoichiometric coefficients (invariant under multiplication), i.e. it should have the same value for the reaction \ce{2H2 + O2 -> 2 H2O} and for the reaction \ce{H2 + $\frac{1}{2}$O2 -> H2O}. Our point of view is that the reaction extent is strongly connected to kinetics, and it is not a tool to describe stoichiometry as some other authors\cite{garst} also think. The only requirement that we accept will be mentioned later, in the discussion of Definition \ref{def:extent}.

We assume throughout that the volume (\(V\)) is constant, and
one can write the generalized form of Eq. \eqref{eq:spec} as
\begin{equation*}
\begin{bmatrix}
N_{\ce{1}}\\
N_{\ce{2}}\\
\dots\\
N_{\ce{M}}
\end{bmatrix}-\begin{bmatrix}
N_{\ce{1}}^0\\
N_{\ce{2}}^0\\
\dots\\
N_{\ce{M}}^0
\end{bmatrix}=
\sum_{r=1}^{R}
\begin{bmatrix}
\gamma_{1,r}\\
\gamma_{2,r}\\
\dots\\
\gamma_{M,r}
\end{bmatrix}W_r, 
\end{equation*}
or shortly
\begin{equation*}
\Nb-\Nb^0=\gammab \Wb,
\end{equation*}
where component \(W_r\) of the vector \(\Wb=\begin{bmatrix}
W_1&W_2&\dots&W_R
\end{bmatrix}\T\) gives the number of occurrences of the \(\rth\) reaction step. 
Note that we do not speak about \emp{infinitesimal} changes. 

With a slight abuse of notation let \(\Wb(t)\), the vector of the numbers of occurrences of reaction events, a step function in the interval \([0,t]\).
Then:
\begin{equation*}
\Nb(t)-\Nb^0=\gammab \Wb(t),
\end{equation*}
or turning to moles
\begin{equation}\label{def:moles}
\nb(t)-\nb^0=\frac{\Nb(t)-\Nb^0}{L}=\gammab \frac{\Wb(t)}{L}=\gammab\xib(t),
\end{equation}
where \(L\)\emp{ is the Avogadro constant} having the unit \({\mol}^{-1}\), and
\begin{equation*}
\nb(t):=\frac{\Nb(t)}{L},
\nb^0:=\frac{\Nb^0}{L},
\xib(t):=\frac{\Wb(t)}{L}.
\end{equation*}
Here we had to choose the less often used notation \(L\) (\url{https://goldbook.iupac.org/terms/view/A00543}) to avoid mixing up with other notations.

The relationship \eqref{def:moles} can be expressed in concentrations as
\begin{equation}
\cb(t)-\cb^0=\frac{\nb(t)-\nb^0}{V}=\frac{\gammab}{V}\xib(t),\label{eq:implicit}
\end{equation}
where \(V\in\R^+\), the volume of the reaction vessel is assumed to be constant, \(\cb(t):=\frac{\nb(t)}{V}\) and
\(\cb^0:=\frac{\nb^0}{V}\). 
The component \(c_{\ce{X($m$)}}\) or \(c_m\) of \(\cb\) is traditionally denoted in chemical textbooks as \([\ce{X($m$)}],\) see e.g. Section 1.2 of Ref.\cite{pillingseakins}.

The concentration in Eq. \eqref{eq:implicit} is again a step function; however, if the number of particles (molecules, radicals, electrons, etc.) is very large, as very often it is, it may be considered to be a continuous, even differentiable function. Remember that the components of \(\xib(t)\) have the dimension of the amount of substance, measured in moles.

Let us now give a general, formal, and \emp{explicit} definition of reaction extent valid \emp{for an arbitrary number of species and reaction steps, and not restricted to mass action type kinetics}. 
(Few qualitative---mainly technical---restrictions are usually made on the function 
\(\rateb\)\cite{feinbergbook,tothnagypapp,volperthudyaev},
but we now mention the continuous differentiability only.)
We start with rewriting the induced kinetic differential equation \begin{equation}\label{eq:ikdegen}
\dot{\cb}(t)=\gammab\rateb(\cb(t))
\end{equation} 
together with the initial condition \(\cb(0)=\cb^0\) into an (equivalent) integral equation:
\begin{equation}\label{eq:int}
\cb(t)-\cb^0=\gammab\int_0^t\rateb(\cb(\overline{t}))
{\;\mathrm{d}\overline{t}}.
\end{equation}

The component \(rate_r\) of the vector \(\rateb\) provides the reaction rate of the \(\rth\) reaction step.
Note that in the mass action case Eq. \eqref{eq:ikdegen} specializes into
\begin{equation}\label{eq:ikdemass}
\dot{\cb}=\gammab\kb\odot\cb^{\alphab} 
\end{equation} 
or, in coordinates
\begin{equation*}
\dot{c}_m(t)=\sum_{r=1}^{R}\gamma_{mr}k_r\prod_{p=1}^{M}c_p^{\alpha_{p,r}}\quad(m=1,2,\cdots,M),
\end{equation*}
where \(\kb\) is the vector of (positive) reaction rate coefficients \(k_r.\)
(We prefer using the expression \emp{reaction rate coefficients} to \emp{reaction rate constants},
as these numbers do depend on many factors---except species concentrations.)
In Eq. \eqref{eq:ikdemass} we used the usual vectorial operations, see e.g. Section 13.2 in T\'oth et al.\cite{tothnagypapp}
Their use in formal reaction kinetics has been initiated by Horn and Jackson.\cite{hornjackson}

In accordance with what has been said up to now, we can introduce the explicit definition of reaction extent by combining Eqs. \eqref{eq:implicit} and \eqref{eq:int}.

\begin{definition}\label{def:extent}
The \emp{reaction extent} 
of a complex chemical reaction or reaction network defined by Eq. \eqref{eq:ccr} is the scalar variable, vector-valued function given by the formula
\begin{equation}\label{eq:specre}
\boxed{
\xib(t):=V\int_0^t
\rateb(\cb(\overline{t}))
{\;\mathrm{d}\overline{t}}}.
\end{equation}
Its time derivative \(\dot{\xib}(t)=V\rateb(\cb(t))\) is usually called the \emp{rate of conversion} or \emp{reaction flux}\cite{polettiniesposito}.
\end{definition}
Note that Eq. \eqref{eq:specre} shows that the reaction extent, in general, depends on the whole history
(past and present) of the vector of concentrations, as if it had a \emp{memory}.\label{page:memory}

Definition \ref{def:extent} of the reaction extent has been derived from the number of reaction events in order to reveal its connection to changes in the concentrations. 
Assuming here also that \(V\) is constant, 
one can formulate the following trivial (equivalent) consequences of the definition:
\begin{equation}\label{eq:trivicons}
\dot{\nb}=\gammab\dot{\xib},\quad\dot{\cb}=\frac{1}{V}\gammab\dot{\xib},\quad
\cb=\cb^0+\gammab\frac{\xib}{V}
\end{equation}
mentioned also by Laidler\cite{laidlerglossary},
sometimes as definitions, sometimes as statements.

Note that neither the rate of the reaction: 
\(\rateb(\cb(t))=\frac{\dot{\xib}(t)}{V},\)
nor the reaction extent \({\xib},\) nor the rate of conversion \(\dot{\xib}\) depends on the stoichiometric matrix \(\gammab\), thereby this
one of the requirements formulated by Dumont et al.\cite{dumonlichanotpoquet}
is fulfilled.

What is wrong with the almost ubiquitous implicit "definition" Eq. \eqref{eq:implicit}? 
We show an example to enlighten this.
\begin{example}
Consider the reaction steps
\begin{equation*}
\ce{X ->[$k_1$] Y}, \quad\ce{X + Y ->[$k_2$] 2Y},
\end{equation*}
expressing the fact that \ce{X} is transformed into \ce{Y} directly and also via autocatalysis.
Although the reaction steps 
\begin{equation*}
\ce{X ->[$k_1$] Y + P},\quad \ce{X + Y ->[$k_2$] 2Y + P}
\end{equation*} with the external species \ce{P}  is a more realistic description of genuine chemical reactions,
e.g. the acid autocatalysis in ester hydrolysis\cite{ostwald,bansagitaylor}, they lead to the same kinetic differential equations for \ce{X} and \ce{Y}. Therefore, we shall analyze the simpler scheme.
Now the stoichiometric matrix $\gamma$ is as follows:
\begin{equation*}
\gammab=\begin{bmatrix}
-1&-1\\
1&1
\end{bmatrix}.
\end{equation*}
Then Eq. \eqref{eq:int} specializes into 
\begin{align*}
c_{\ce{X}}(t)-c_{\ce{X}}(0)&=
-\int_{0}^{t}k_1c_{\ce{X}}(\overline{t}){\;\mathrm{d}\overline{t}}
-\int_{0}^{t}k_2c_{\ce{X}}(\overline{t})c_{\ce{Y}}(\overline{t}){\;\mathrm{d}\overline{t}}
=
-\frac{\xi_1(t)+\xi_2(t)}{V}\\
c_{\ce{Y}}(t)-c_{\ce{Y}}(0)&=
\int_{0}^{t}k_1c_{\ce{X}}(\overline{t}){\;\mathrm{d}\overline{t}}+
\int_{0}^{t}k_2c_{\ce{X}}(\overline{t})c_{\ce{Y}}(\overline{t}){\;\mathrm{d}\overline{t}}
=
\frac{\xi_1(t)+\xi_2(t)}{V}.
\end{align*}
These two relations do not determine \(\xi_1\) and \(\xi_2\) individually but only their sum (even if one utilizes \(c_{\ce{Y}}(t)=c_{\ce{X}}(0)+c_{\ce{Y}}(0)-c_{\ce{X}}(t)\)).
The problem originates from the fact that the reaction steps are not linearly independent as reflected in the singularity of the matrix \(\gammab.\)
\end{example}

If the reaction steps of a complex chemical reaction are independent, the situation is better.
\begin{example}\label{ex:indep}
In some special cases, there is a way of making the "definition" Eq. \eqref{eq:implicit} into a real, explicit definition.
Assume that \(R\le M\), and that the stoichiometric matrix \(\gammab\) is of the full rank, i.e. the reaction steps are independent. 
Then, one can rewrite Eq. \eqref{eq:implicit} in two steps as follows:
\begin{align}
\gammab\T(\cb(t)-\cb^0)&=\frac{1}{V}\gammab\T\gammab\xib(t)\nonumber\\
\xib(t)&=V(\gammab\T\gammab)^{-1}\gammab\T(\cb(t)-\cb^0).\label{eq:memoryless}
\end{align}
Now one can accept Eq. \eqref{eq:memoryless} as a definition for the reaction extent.
Nevertheless, in this special case 
Eq. \eqref{eq:int} 
implies
\begin{equation*}
\gammab\T(\cb(t)-\cb^0)=\gammab\T\gammab\int_0^t\rateb(\cb(\overline{t})){\;\mathrm{d}\overline{t}}
\end{equation*}
and
\begin{equation*}
(\gammab\T\gammab)^{-1}\gammab\T(\cb(t)-\cb^0)=
\frac{1}{V}\xib(t)=
\int_0^t\rateb(\cb(\overline{t})){\;\mathrm{d}\overline{t}},
\end{equation*}
thus this definition is the same as the one in Eq. \eqref{eq:specre}. 
This derivation can always be done if \(R=1\) that is, in a not-so-interesting trivial case.
Unfortunately, the case \(R\le M\) does not happen very often. On the contrary, for example, in
case of combustion reactions, Law's law\cite{law} (see page 11) states that \(R\approx 5 M.\)

Note also, that Eq. \eqref{eq:memoryless} shows the  following:  in these cases, i.e. when the stoichiometric matrix is of the full rank---as opposed to the general case, see page \pageref{page:memory}---the reaction extents do not depend on the whole history of the concentration vector, it only depends on its instantaneous value.
\end{example}
Let us make a trivial remark on the independence of reaction steps. 
If the complex chemical reaction consists of a single irreversible step, then the reaction steps(!) are independent. 
If any of the reaction steps are reversible, then the reaction steps are not independent. 
\subsection{Properties of the reaction extent}
The usual assumptions \label{pg:assump} on the vector-valued function \(\rateb\) are as follows, see Refs.\cite{tothnagypapp,volperthudyaev}.
\begin{enumerate}
\item
All of its components are continuously differentiable functions defined on \(\R^M\) taking only non-negative values. This is usual, e.g. in the case of mass action kinetics, but---with some restrictions---also in the case when the reaction rates are rational functions as in the case of Michaelis--Menten or Holling type kinetics, see e.g. Refs.\cite{polczkulcsarszederkenyi,polczpeniszederkenyi,kisstoth,laidlerglossary}
\item
The value of \(rate_r(\cb)\)  is zero if and only if some of the species needed for the \(\rth\) reaction step is missing, i.e. for some \(m: \alpha_{m,r}>0\) and \(c_m=0\) (see p. 613, Condition 1 in Ref.\cite{volperthudyaev}). 
We shall say in this case that reaction step \(r\) \emp{cannot start} from the concentration vector  \(\cb.\)
\end{enumerate}
The second assumption implies---even in the general case, i.e. without restriction to the mass action type kinetics---that \(rate_r(\cb)>0\) if all the necessary species (\emp{reactants}, see below) are present initially: \(\alpha_{m,r}>0\Longrightarrow c_m>0.\)

Let us sum up the relevant qualitative characteristics of the reaction extent.
(Remember that the proof can be found in the Appendix.)
\begin{theorem}\label{thm:basics}
\begin{enumerate}
\item[\nonumber]
\item
The domain of the function \(t\mapsto\xib(t)\) is the same as that of \(\cb\).
\item
Both \(\cb\) and \(\xib\in\CC^2(J,\R^R);\) with some open interval \(J\subset\R\) such that \(0\in J.\)
 \item
 \(\xib\) obeys the following initial value problem:
\begin{equation}\label{eq:rmdiffegy}
\dot{\xib}(t)=V\rateb(\cb^0+\frac{1}{V}\gammab\xib(t)),\quad \xib(0)=\nulb.
\end{equation}
\item
At the beginning, the velocity vector of the reaction extent (also called the \emp{rate of conversion}) points into the closed first orthant, and this property is kept for all times in the domain of the function \(t\mapsto\xib(t).\) 
\item
The components of \(\xib\) are either positive, strictly monotonously increasing functions or constant zero. 
If for some positive time \(t\) we find that \(\xi_r(t)=0\) then, obviously, the reaction step \(r\) did not start at all at the beginning.
\end{enumerate}
\end{theorem}

Let us make a few remarks:
\begin{itemize}
\item 
The last property (positivity) mentioned in the Theorem can be realized with \(\lim_{t\to+\infty}\xi(t)=+\infty\)
(the simplest example for this being \(\ce{X -> 2X}, c_{\ce{X}}^0>0\)),  or with a finite positive value of \(\lim_{t\to+\infty}\xi(t),\) see the example \ce{X -> Y -> Z} below.
\item
Eq. \eqref{eq:rmdiffegy} shows that we would have got simpler formulas if we used
\(\frac{\xi(t)}{V}\) as proposed by Aris\cite{arisintro} on p. 44, 
but this form is valid only if \(V\) is constant. 
\item
In the mass action case both \(\cb\) and \(\xib\) are infinitely many times differentiable.
\item
If one uses a kinetics different from the mass action type
not fulfilling assumptions 1 and 2 on page \pageref{pg:assump}, 
or---as P\'ota\cite{potarabai} has shown---if one applies an approximation, 
then it may happen that some of the initially positive concentrations turn to zero.
\end{itemize}

In order to proceed, we need to make a technical remark on the figures shown hereinafter. We label the first axis (usually: horizontal) in the figures with \(t\mathrm{/s}\), where \(\mathrm{s}\) is the time unit second. Labels of other axes are formed in a similar way. With this procedure we want to emphasize that the figures show the relationship between pure numbers and not between physical quantities. 

The condition in part 5 of Theorem \ref{thm:basics} is only necessary but not sufficient as the example below shows.  
\begin{example}\label{ex:convexconcave}
Let us start the consecutive reaction \ce{X ->[$k_1$] Y ->[$k_2$] Z} from the vector of the initial concentrations:
\(\begin{bmatrix}
c^0_{\ce{X}}&0&0
\end{bmatrix}\T\), and suppose \(k_1\neq k_2.\) 
Although, the second step cannot start at the beginning, yet the second reaction extent is positive for all positive times as the solution of the evolution equations
\begin{equation}\label{eq:velo}
\dot{\xi}_1=Vk_1\left(c^0_{\ce{X}}-\frac{\xi_1}{V}\right),\quad
\dot{\xi}_2=Vk_2\left(0+\frac{\xi_1-\xi_2}{V}\right)
\end{equation}
are as follows
\begin{align*}
&\xi_1(t)=Vc^0_{\ce{X}}(1-e^{-k_1t}),\\
&\xi_2(t)=\frac{Vc^0_{\ce{X}}}{k_2-k_1}\left(k_2(1-e^{-k_1t})-k_1(1-e^{-k_2t})\right).
\end{align*}
Positivity also follows without any calculations from the fact that the velocity vector of the differential equations in \eqref{eq:velo} point inward, into the interior of the first quadrant, or using the fact that Eqs. \eqref{eq:velo} are also kinetic type differential equations.

Note that \(\lim_{t\to+\infty}\xi_1(t)=\lim_{t\to+\infty}\xi_2(t)=Vc^0_{\ce{X}}.\)
It means that the number of occurrences of the reaction events for both reactions, and thus the reaction extents, are exactly the same at the end of the whole process. 
Moreover, it does not depend on the reaction rate coefficients.

Easy calculations show the following facts. 
The function \(\xi_2\,/\mol\) in Fig. \ref{fig:convexconcave} has an inflection point, because its second derivative is zero at some positive time \(t_{\mathrm{infl}}\) for all choices of the reaction rate coefficients, and the third derivative is not zero at \(t_{\mathrm{infl}}\). 
The function \(\xi_1\,/\mol\) in Fig. \ref{fig:convexconcave} is concave from below no matter what the reaction rate coefficients are. 
\begin{figure}[!ht]
\centering
\includegraphics[width=0.8\linewidth]{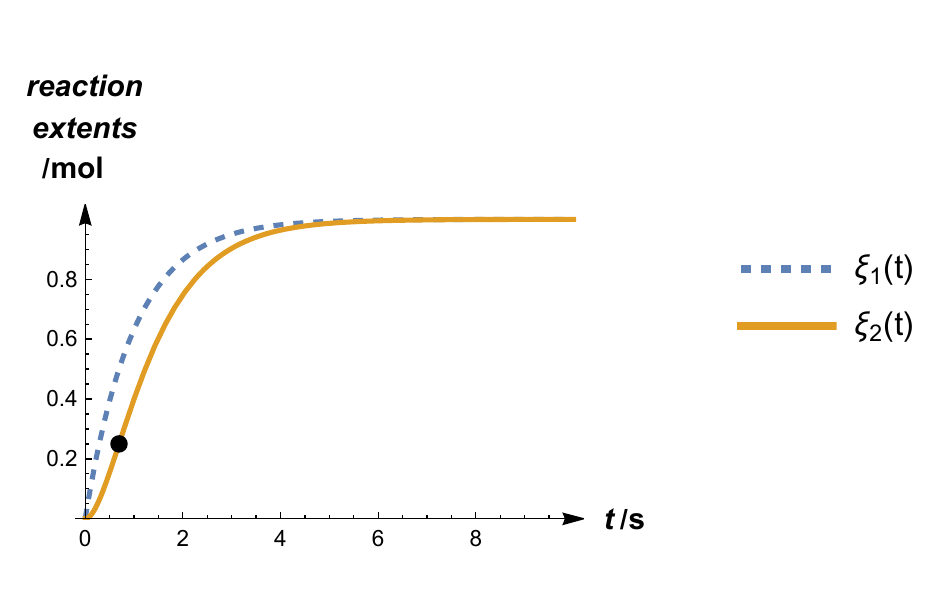}
\caption{Reaction extents in the consecutive reaction \ce{X ->[$k_1$] Y ->[$k_2$] Z} when 
\(k_1~=~1~{\mathrm{s}}^{-1}\), 
\(k_2=~2~{\mathrm{s}}^{-1}\), 
\(c_{X}^0=~1~\mL\), 
\(c_{\ce{Y}}^0=c_{\ce{Z}}^0=~0~\mL\), 
\(V=~1~\Lit.\) 
The limit 1 here is only a consequence of the choice of the parameters. 
The large dot is the inflection point of the curve \(\xi_2\,/\mol\).}
\label{fig:convexconcave}
\end{figure}
\end{example}

To characterize the convexity of the reaction extents in the general case is an open problem. 
One should take into consideration that although in the practically interesting cases the number of equations in \eqref{eq:rmdiffegy} is larger than those in Eq.
\eqref{eq:ikdegen}, that is \(R > M\), the  equations for the reaction extents are of a simpler structure. 

Monotonicity mentioned in Theorem \ref{thm:basics} implies that all the components of the reaction extent do have a finite or infinite limit as \(t\) tends to \(\sup(J)=:t^*,\) where \(J:=\Dom(\cb),\)
and \(t^*\) is a finite or infinite time. 
It is an interesting open question: 
when does a coordinate of the reaction extent vector tend to infinity?
\begin{figure}[!ht]
\centering
\includegraphics[width=0.9\linewidth]{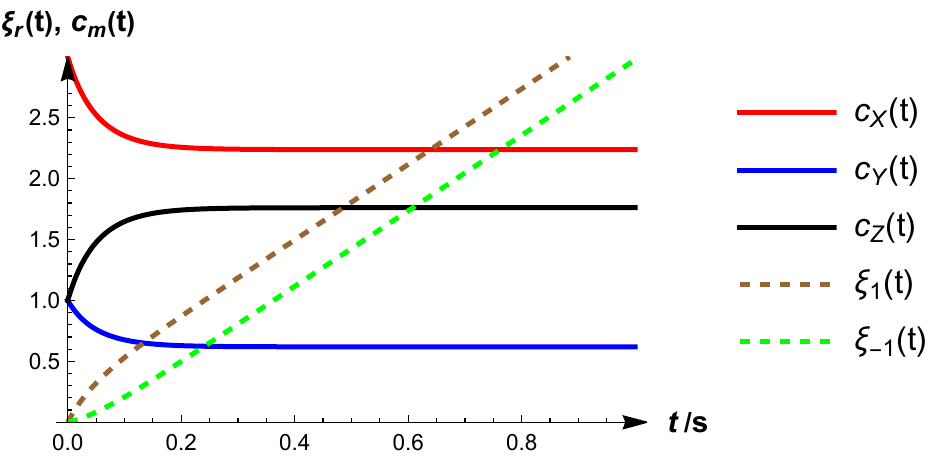}
\caption{
\(c_{\ce{X}}^0~=~3.0~\mL\),
\(c_{\ce{Y}}^0~=~1.0~\mL\),
\(c_{\ce{Z}}^0~=~1.0~\mL\),
\(k_1~=~1.0~\Litsix~\mol^{-2}~\mathrm{s}^{-1}\),
\(k_{-1}~=~1.0~\Lit~\mol^{-1}~\mathrm{s}^{-1}\),
\(V~=~1~\Lit.\)
While the concentrations tend to and are becoming very close to the equilibrium values, the reaction extents tend to infinity in a monotonously increasing way in the reaction: 
\ce{2X + Y <=>[$k_1$][$k_{-1}$] 2Z}.}
\label{fig:dynamiceq}
\end{figure}
As the \emp{emphatic} closure of this series of remarks, we mention that the strictly monotonous increase of the number of the occurrence of reaction events shows that the reaction events never stop, see \ref{fig:dynamiceq}. This important fact is independent on the form of kinetics, and it is a property of the deterministic models of reaction kinetics.
This sheds light on the meaning of \emp{dynamic equilibrium} as generally taught. 
Note that \emp{no reference to thermodynamics or statistical physics} has been invoked here,
analyzing the connections are left to the reader.
\begin{example}\label{ex:blowup}
In case when the domain of the function \(t\mapsto\cb(t)\) is a proper subset of the non-negative real numbers, \(\xib\) has the same property. 
Let us consider the \ikde\ \(\dot{c}=kc^2\) of the quadratic auto-catalytic reaction \ce{2X ->[k] 3X} with the initial condition \(c(0)=c^0>0.\) Then, 
\(
c(t)=\frac{c^0}{1-kc^0t},  \left(t\in\left[0,\frac{1}{kc^0}\right[
\,\subsetneq\,[0,+\infty[\right).
\) 
Now Eq. \eqref{eq:rmdiffegy} specializes into
\(\dot{\xi}=Vk(c^0+\xi/V)^2,{\ }\xi(0)=0\)
having the solution
\(
\xi(t)=V(c^0)^2\frac{kt}{1-kc^0t},  \left(t\in\left[0,\frac{1}{kc^0}\right[\right),
\)
thus \(\xi\) blows up at the same time (\(t^*:=\frac{1}{kc^0}\)) when \(c\) does.
Up to the blow-up, the reaction event occurs infinitely many times:
\(\lim_{t\to+t^*}\xi(t)=+\infty.\)
Definitions and a few statements about blow-up in kinetic differential equations are given in the works by Csikja et al.\cite{csikjapapptoth,csikjatoth}
\end{example}

\section{What is it that tends to 1?}\label{sec:what}
Our interest up to this point was the number of occurrences of reaction events. 
However, many authors think it is useful and visually attractive that the "reaction extent tends to 1 when the reaction tends to its end," see e.g. Fig. 1 of Glasser\cite{glasser}.
Borge\cite{borge} and Peckham\cite{peckham} also argue for [0,1].

Another approach is given by Moretti\cite{moretti}, Dumon et al.\cite{dumonlichanotpoquet} and others via introducing the \emp{reaction advancement ratio} \(\frac{\xi}{\xi_{\max}}\), and stating that this ratio is always between 0 and 1. 
Peckham\cite{peckham} noticed that Atkins\cite{atkins5} (pp. 272--276) shows a figure of the free energy \(G\) of the reacting system versus \(\xi,\) where the first axis is labeled from zero to one. 
However, in the next edition\cite{atkins6} (pp. 216--217), the graph has been changed, and it now shows the first axis without 1 as an upper bound.
Being loyal to the usual belief\cite{peckham,treptow}, 
we are looking for quantities (pure numbers) tending to 1 as e.g. \(t\to+\infty\). 
Scaling might help to find such quantities.
\subsection{Scaling by the initial concentration: One reaction step}
Now we are descending from the height of generality by considering a single irreversible reaction step (\(R=1\)), assuming that the kinetics is of the mass action type. Thus, the reaction is
\begin{equation}\label{eq:onestep}
\ce{$\sum_{m=1}^M\alpha_m$X($m$) ->[k] $\sum_{m=1}^M\beta_m$X($m$)}.
\end{equation}
Therefore, one has the reaction extent
\begin{equation}\label{eq:ximone}
\dot{\xi}=Vk\left(\begin{bmatrix}
c_1^0\\c_2^0\\\dots\\c_M^0
\end{bmatrix}+
\begin{bmatrix}
\gamma_1\\\gamma_2\\\dots\\\gamma_M
\end{bmatrix}\frac{\xi}{V}\right)^{\alphab}
=
Vk\prod_{m=1}^{M}\left(c_m^0+\gamma_m\frac{\xi}{V}\right)^{\alpha_m},\quad \xi(0)=0;
\end{equation}
with \(\gamma_m:=\beta_m-\alpha_m.\)

\begin{theorem}\label{thm:singlestep}
\begin{enumerate}
\item[\nonumber]
\item 
If the reaction in Eq. \eqref{eq:onestep} cannot start, then
\(\xi(t)=0\) for all non-negative real times \(t\):
\begin{equation}\label{eq:cannot}
\exists m:(\alpha_m\neq0\ \&\ c_m^0=0)\Longrightarrow\forall t\in\R^+_0:\xi(t)=0.
\end{equation}
\item
If the reaction in Eq. \eqref{eq:onestep} does start and all the species are produced (i.e. for all \(m: \gamma_m>0\)), then \(\xi(t)\)
tends to infinity (blow-up included):
\begin{equation}\label{eq:produced}
\forall m:\gamma_m:=\beta_m-\alpha_m>0\Longrightarrow\lim_{t\to t^*}\xi(t)=+\infty,
\end{equation} 
where \(t^*:=\sup(J)\) with \(J:=\Dom(\xi)\).
\item 
If some of the species is consumed, that is \(\exists m: \gamma_m<0\), then 
\begin{equation}\label{eq:decreases}
\lim_{t\to +\infty}\xi(t)=\min\left\{-\frac{Vc_m^0}{\gamma_m};\gamma_m<0\right\}.
\end{equation}
\end{enumerate}
\end{theorem}

\begin{example}
\begin{enumerate}
\item [\nonumber]
\item 
Reaction \ce{X ->[$k$] Y} with \(c^0_{\ce{X}}=0\) and \(c^0_{\ce{Y}}\)
arbitrary, illustrates the first case as here \(\dot{\xi}=-k\xi, \xi(0)=0\)
implies \(\forall t\in\R:\xi(t)=0.\)
\item 
Reaction \ce{X ->[$k$] 2X} with \(c^0_{\ce{X}}>0\) is an illustration for the second case 
\begin{equation*}
\forall t\in\R:\xi(t)=V c^0_{\ce{X}}(e^{k t}-1 )
\end{equation*}
with 
\(
\lim_{t\to+\infty}\xi(t)=+\infty.   
\)
\item 
Reaction \ce{2X ->[$k$] 3X} with \(c^0_{\ce{X}}>0\) (Example \ref{ex:blowup}) is another illustration for the second case with 
\begin{equation*}
\forall t\in[-\infty,\frac{1}{kc^0_{\ce{X}}}[ :\xi(t)=\frac{k t V (c^0_{\ce{X}})^2}{1-ktc^0_{\ce{X}}}
\end{equation*} with
\(\lim_{t\to\frac{1}{kc^0_{\ce{X}}}}\xi(t)=+\infty.\)
\item 
Reaction \ce{X + Y ->[$k$] 2X}
is an illustration for the third case 
\begin{equation*}
\forall t\in\R:\xi(t)=
\frac{(-1 + e^{k t (c^0_{\ce{X}} + c^0_{\ce{Y}})}) V c^0_{\ce{X}} c^0_{\ce{Y}}}{(c^0_{\ce{Y}} + e^{k t (c^0_{\ce{X}} + c^0_{\ce{Y}})} c^0_{\ce{X}})}
\end{equation*} 
with \(\lim_{t\to+\infty}\xi(t)=Vc^0_{\ce{Y}},\) if \(c^0_{\ce{X}},c^0_{\ce{Y}}\neq0\). If either \(c^0_{\ce{X}}=0,\) or \(c^0_{\ce{Y}}=0,\)
then \(\forall t\in \R:\xi(t)=0.\)
\item
The example \ce{X -> Y} with \(c_{\ce{X}}^0=0, c_{\ce{Y}}^0>0\) shows that
a species (here \ce{Y}) can have positive concentration for all positive times in a reaction where "none of the steps" can start.
\end{enumerate}
\end{example}

The table below shows a series of examples illustrating different types of single irreversible reaction steps.

\begin{table}
  \caption{Reaction extent for various reaction types}
  \label{table:cases}
\begin{tabular}{lllll}
\hline
Step                      &\(\cb^0\)&\(\dot{\xi}=\)                              &\({\xi}(t)=\)                                                &Case\\ 
\hline\ce{X ->[$k$] 2X}     & 0       &\(k\xi\) & 0             &\eqref{eq:cannot}\\
\hline\ce{X ->[$k$] 2X}     & 1       &\(Vk(1+\xi/V)\)& \(V(e^{kt}-1)\)&\eqref{eq:produced}\\
\hline\ce{2X ->[$k$] 3X}    & 1       &\(Vk(1+\xi/V)^2\)                               &\(\frac{Vkt}{1-kt}\)                                           &\eqref{eq:produced}\\
\hline\ce{2X + Y ->[$k$] 2Z}&         &\(Vk(c_{\ce{X}}^0-2\xi/V)^2*\)&\\
&&\((c_{\ce{Y}}^0-\xi/V)\)&            &\eqref{eq:decreases}\\
\hline
\end{tabular}
\end{table}

\begin{figure}[!ht]
\centering
\includegraphics[width=0.7\linewidth]{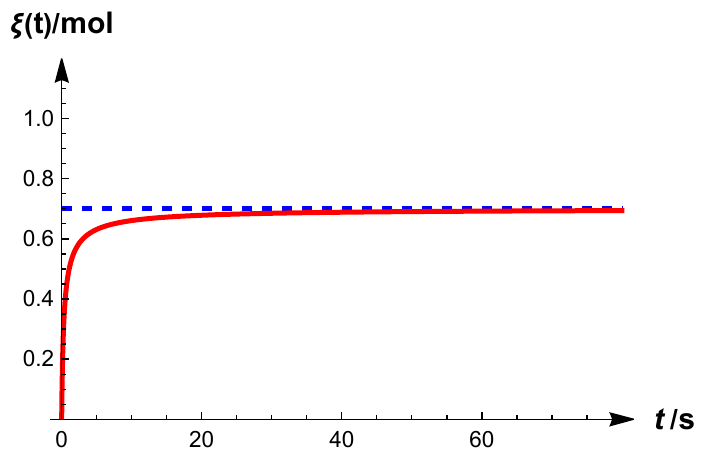}
\caption{
\(c_{\ce{X}}^0/2~=~0.7~\mL<c_{\ce{Y}}^0~=~1.2~\mL\),
\(c_{\ce{Z}}^0~=~3~\mL\),
\(k~=~1.0~\Litsix~\mol^{-2}~\mathrm{s}^{-1}\),
\(V~=~1~\Lit.\)
\ce{Y} is in stoichiometric excess in the reaction \ce{2X + Y -> 2Z}. 
The limiting value is denoted with the dashed line.} 
\label{fig:rm11}\end{figure}

\begin{example}
Here we analyze the last example of Table \ref{table:cases}.
In the case of the reaction \ce{2X + Y ->[$k$] 2Z} mimicking water formation one has the following quantities:
\(R:=1,M:=3,\ce{X}:=\ce{H2}, \ce{Y}:=\ce{O2},\ce{Z}:=\ce{H2O}.\)
Furthermore,
\[
\alphab=
\begin{bmatrix}
2\\1\\0
\end{bmatrix}{\!}{\!},\quad
\betab=\begin{bmatrix}
0\\0\\2
\end{bmatrix}{\!}{\!},\quad 
\gammab=\begin{bmatrix}
-2\\-1\\2
\end{bmatrix}{\!}{\!}.
\]
The initial value problem to describe the time evolution of reaction extent is
\begin{equation}\label{eq:wateru}
\dot{\xi}=Vk\left(
\begin{bmatrix}
c_{\ce{X}}^0\\c_{\ce{Y}}^0\\c_{\ce{Z}}^0
\end{bmatrix}+
\begin{bmatrix}
-2\\-1\\2\\
\end{bmatrix}\frac{\xi}{V}\right)^{\begin{bmatrix}
2\\1\\0
\end{bmatrix}}
=
Vk(c_{\ce{X}}^0-2\frac{\xi}{V})^2(c_{\ce{Y}}^0-\frac{\xi}{V}),\quad \xi(0)=0.
\end{equation}
We can provide only the inverse of the solution to Eq. \eqref{eq:wateru}. However, one can state that the reaction extent tends strictly monotonously to its limit (independent on the value of the reaction rate coefficient): \(\lim_{t\to+\infty}\xi(t)=\min\{\frac{Vc_{\ce{X}}^0}{2},Vc_{\ce{Y}}^0\}.\)
Different initial conditions lead to different results, see Figs. \ref{fig:rm11}--\ref{fig:rm13}. Obviously,  the third point of Theorem \ref{thm:singlestep} is of main practical use here. For this case one has the following statement.
\end{example}

\begin{corollary}\label{corr:pure}
In the third case of Theorem \ref{thm:singlestep}, 
dividing \(\xi\) by the initial concentration and scaled by the quantity \(-\gamma_m/V\), 
we obtain a (pure) number tending to 1 as \(t\) tends to infinity:
\fbox{\(\lim_{t\to +\infty}
\frac{\xi(t)}{Vc_m^0/(-\gamma_m)}=1.\)}
\end{corollary}
\subsection{Stoichiometric initial condition, excess and deficit}
Before studying the above-mentioned figures, we need some definitions in order to avoid the sin of using a concept without having defined it. The concepts of \emp{stoichiometric initial condition} and \emp{initial stoichiometric excess} are often used elsewhere but never defined.
\begin{definition}
Consider the induced kinetic differential equation 
\eqref{eq:ikdegen} of the reaction \eqref{eq:ccr} with the initial condition
\(\cb(0)=\cb^0\neq\nulb.\)
This initial condition is said to be a \emp{stoichiometric initial condition} (and \(\cb^0\) is a stoichiometric initial concentration), if for all such
\(m=1,2,\dots,M;{\ }r=1,2,\dots,R\) for which \(\gamma_{m,r}<0\) the ratios \(\frac{c_m^0}{-\gamma_{m,r}}\)
are independent from \(m\) and \(r.\) 
If the ratios are independent of \(r\), but for some 
\(p=1,2,\dots,M\) the ratio \(\frac{c_p^0}{-\gamma_{p,r}}\) is larger than the others, then \ce{X(p)} is said to be in \emp{initial stoichiometric excess}, or it is in excess initially.
If the ratios are independent of \(r\), but for some 
\(p=1,2,\dots,M\) the ratio \(\frac{c_p^0}{-\gamma_{p,r}}\) is smaller than the others, then \ce{X(p)} is said to be in \emp{initial stoichiometric deficit}, or it is in deficit initially.
The last notion is mathematically valid, but in such cases, one prefers saying that all the other species are in excess.
In combustion theory the expressions \emp{stoichiometric}, \emp{fuel lean} and \emp{fuel rich} are used in the same sense, see page 115 of the book by Tur\'anyi and Tomlin\cite{turanyitomlin}.
\end{definition}

\begin{figure}[!ht]
\centering
\includegraphics[width=0.7\linewidth]{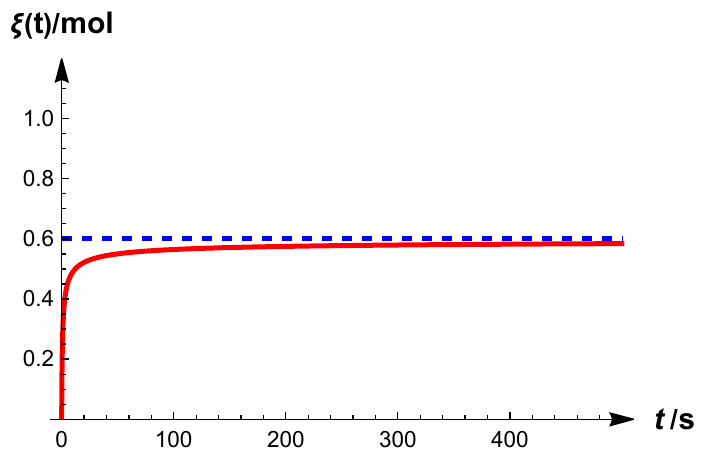}
\caption{
\(c_{\ce{X}}^0/2~=c_{\ce{Y}}^0=~0.6~\mL\),
\(c_{\ce{Z}}^0~=~3~\mL\), 
\(k~=~1.0~\Litsix~\mol^{-2}~\mathrm{s}^{-1}\),
\(V~=~1~\Lit.\) Stoichiometric initial condition in the reaction \ce{2X + Y -> 2Z}; slow convergence. The limiting value is denoted with the dashed line.}
\end{figure}

\begin{figure}[!ht]
\centering
\includegraphics[width=0.7\linewidth]{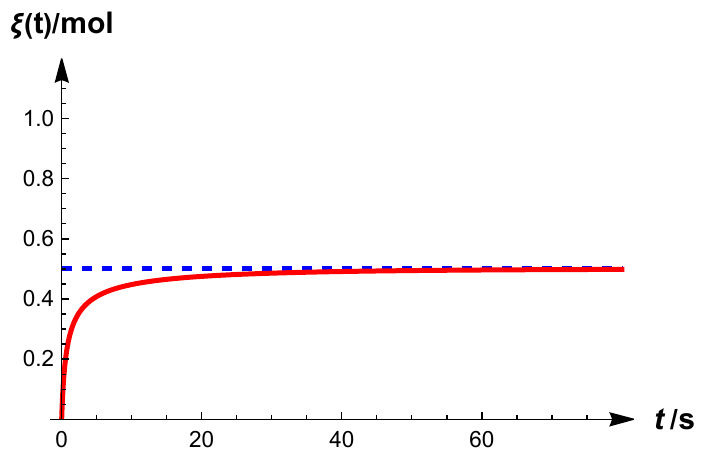}
\caption{
\(c_{\ce{X}}^0/2~=~0.6~\mL>c_{\ce{Y}}^0~=~0.5~\mL\),
\(c_{\ce{Z}}^0~=~3~\mL\), 
\(k~=~1.0~\Litsix~\mol^{-2}~\mathrm{s}^{-1}\),
\(V~=~1~\Lit.\)
\ce{X} is in stoichiometric excess in the reaction \ce{2X + Y -> 2Z}. 
The limiting value is denoted with the dashed line.}
\label{fig:rm13}\end{figure}

\begin{figure*}
\centering
\includegraphics[width=0.3\linewidth]{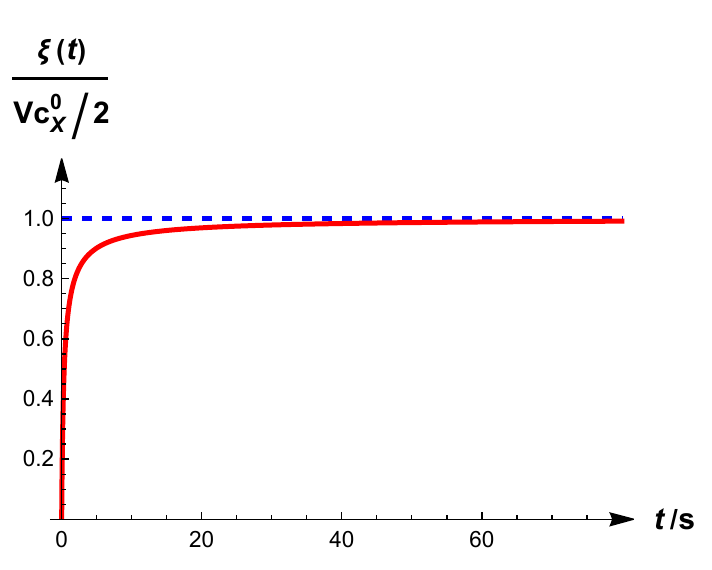}\quad
\includegraphics[width=0.3\linewidth]{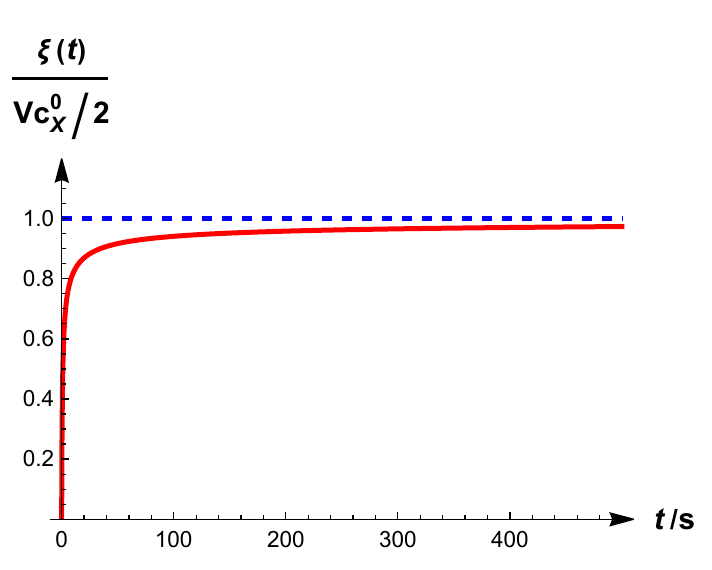}\quad
\includegraphics[width=0.3\linewidth]{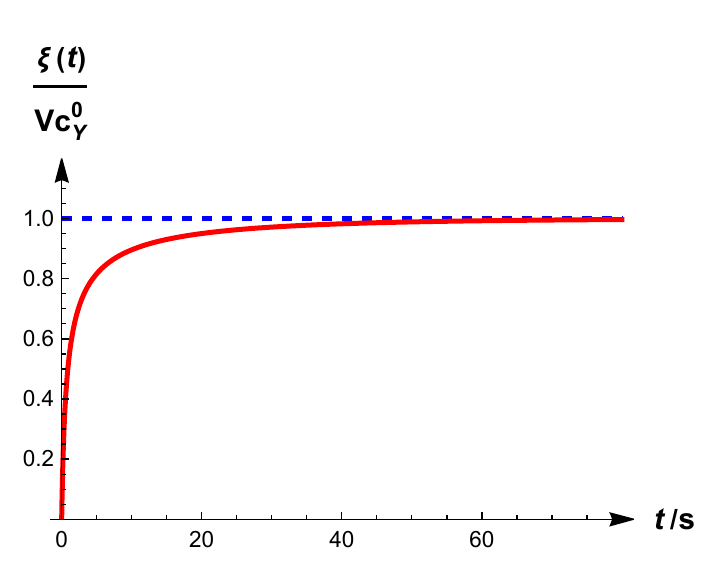}
\caption{Scaled reaction extents tend to 1. 
The reaction rate coefficient and the initial data are the same as in Figs. \ref{fig:rm11}--\ref{fig:rm13}.}
\label{fig:rm1skalazott}
\end{figure*}
We suggest that instead of saying that the scaling factor is some initial concentration as in the third case of Theorem \ref{thm:singlestep},
one can equally well say that the divisor is the limiting value of the reaction extent, as it is in the cases in figures \ref{fig:rm1skalazott}: \(Vc_{\ce{X}}^0/2, Vc_{\ce{X}}^0/2=Vc_{\ce{Y}}^0,
Vc_{\ce{Y}}^0,\) respectively.
This result will come in handy below.

Under stoichiometric initial conditions Peckham\cite{peckham} gives a definition of \(\xi_{\max}\) and proposes to use \(\frac{\xi(t)}{\xi_{\max}}\) in extremely special cases. The domain of this ratio is \([0,1].\)

At this point it may not be obvious how to generalize Corollary \ref{corr:pure}. In order to treat more complicated cases we shall choose another way.
\subsection{Scaling by the "maximum"}
In cases when \(\xi^*:=\lim_{t\to+\infty}\xi(t)\) is finite, then 
\(\xi^*=\sup\{\xi(t);t\in\R\}\), thus  
 \(\xi^*\) may be identified (mathematically incorrectly) with \(\xi_{\max},\) and surely \(\lim_{t\to+\infty}\frac{\xi(t)}{\xi^*}=1.\)
 That is the procedure applied by most authors\cite{dumonlichanotpoquet,vandezandevandergrienddekock,moretti}.
\subsection{Detailed balanced reactions}
\begin{definition}
The complex chemical reaction
\begin{equation}\label{eq:revccr}
\ce{$\sum_{m=1}^M\alpha_{m,r}$X($m$) <=>[$k_r$][$k_{-r}$] $\sum_{m=1}^M\beta_{m,r}$X($m$)}\quad (r=1,2,\dots,R);
\end{equation}
endowed with mass action kinetics is said to be \emp{conditionally detailed balanced} at the positive stationary point \(\cb^*\) if
\begin{equation}\label{eq:dbgeneralcond}
k_r(\cb^*)^{\alphab_{.,r}}=k_{-r}(\cb^*)^{\betab_{.,r}}
\end{equation}
holds. It is unconditionally \emp{detailed balanced} if
Eq. \eqref{eq:dbgeneralcond} holds for any choice of (positive) reaction rate coefficients.
\end{definition}
Note that all the steps in \eqref{eq:revccr} are reversible. Furthermore, in such cases the reaction steps are indexed
by \(r\) and \(-r,\). It is always our choice in which order the forward and backward steps are written, expressing the fact that "forward" and "backward" has no true physical meaning.
\subsubsection{Ratio of two reaction extents.}
Suppose we have a reversible reaction
\begin{equation}\label{eq:singledb}
\ce{$\sum_{m=1}^M\alpha_m$X($m$) <=>[$k_1$][$k_{-1}$] $\sum_{m=1}^M\beta_m$X($m$)}
\end{equation}
being unconditionally detailed balanced because the number of the forward and backward reaction pairs is 1.
Then the initial value problem for the reaction extents is as follows.
\begin{align*}
&\dot{\xi}_1=Vk_1\prod_{m=1}^{M}(c_m^0+\gamma_m(\xi_1-\xi_{-1})/V)^{\alpha_m},&\xi_1(0)=0,\\
&\dot{\xi}_{-1}=Vk_{-1}\prod_{m=1}^{M}(c_m^0+\gamma_m(\xi_1-\xi_{-1})/V)^{\beta_m},&\xi_{-1}(0)=0,
\end{align*}
where \(\gamma_m:=\beta_m-\alpha_m.\)
\begin{proposition}\label{prop:specdb}
Under the above conditions, one has
\fbox{\(\lim_{t \to +\infty}
\frac{\xi_1(t)}{\xi_{-1}(t)}=1.\)}
\end{proposition}
Note that initially one only knows that it is the \emp{derivative}s of the reaction extents that have the same value at equilibrium.

\begin{example}\label{ex:waterform}
Consider the reversible reaction
\ce{2X + Y <=>[$k_1$][$k_{-1}$] 2Z} for water formation with the data 
\( k_1~=~1~\Litsix~\mol^{-2}~\mathrm{s}^{-1},\) 
\(k_{-1}~=~1~\Lit\mol^{-1}~\mathrm{s}^{-1},\)
\(c_{\ce{Y}}^0~=~1~\mL, c_{\ce{Z}}^0~=~1~\mL.\)
\begin{itemize}
\item 
If \(c_{\ce{X}}^0~=~3~\mL, \) then \ce{X} is in excess initially (a);
\item
if \(c_{\ce{X}}^0~=~2~\mL, \) then one has a stoichiometric initial condition (b);
\item
if \(c_{\ce{X}}^0~=~1~\mL, \) or
\(c_{\ce{X}}^0~=~1/2~\mL,\) then \ce{Y} is in excess initially (c or d).
\end{itemize}
Note that it is not the excess or deficit that is relevant, see Conjecture \ref{conj:iniratio} below.
The initial rates of the forward and backward reactions are as follows:
\begin{itemize}
\item 
\(1\cdot9\cdot1 > 1\cdot1,\)
\item 
\(1\cdot4\cdot1 > 1\cdot1,\)
\item 
\(1\cdot1\cdot1 = 1\cdot1,\)
\item 
\(1\cdot1/4\cdot1 <1\cdot1.\)
\end{itemize}
The results are in accordance with Conjecture \ref{conj:iniratio} below and can be seen in Figs. \ref{fig:rm21} and \ref{fig:rm22}.
\begin{figure}
\centering
\includegraphics[width=0.7\linewidth]{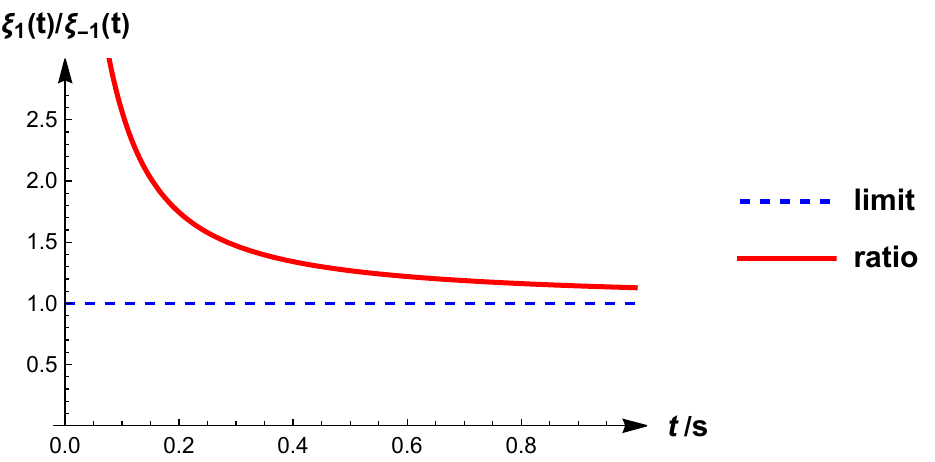}\\
\includegraphics[width=0.7\linewidth]{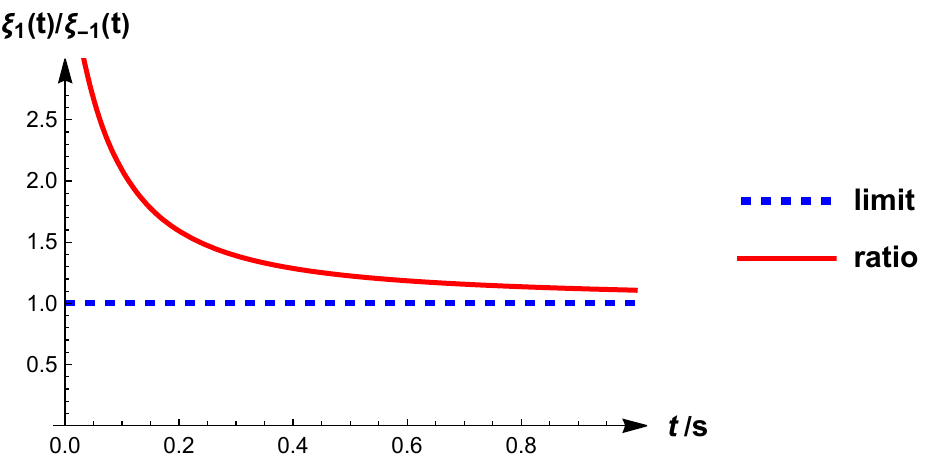}\\
\caption{The ratio \(\frac{\xi_{1}(t)}{\xi_{-1}(t)}\) is tending to 1 from above: \ce{X} is in excess at the top figure (case a of Example \ref{ex:waterform}) and the initial condition is stoichiometric at the bottom figure (case b) in case of the reaction \ce{2X + Y <=>[$k_1$][$k_{-1}$] 2Z}. \(V~=~1~\Lit,\) and other data are given in the text.}
\label{fig:rm21}
\end{figure}
\begin{figure}
\centering
\includegraphics[width=0.7\linewidth]{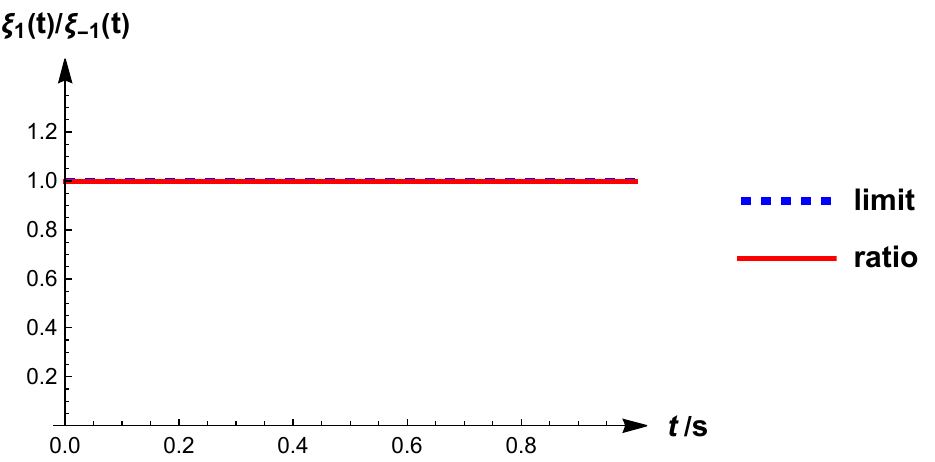}\\
\includegraphics[width=0.7\linewidth]{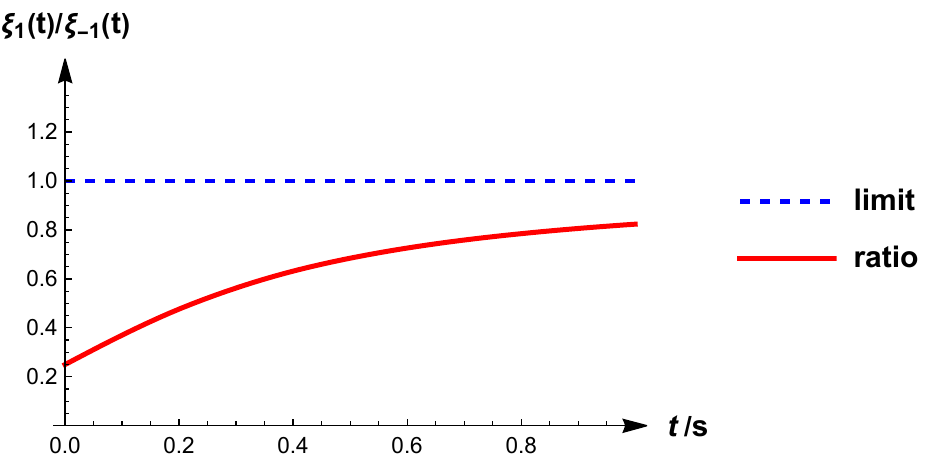}\\
\caption{The ratio \(\frac{\xi_{1}(t)}{\xi_{-1}(t)}\) is constant at the top figure (case c of Example \ref{ex:waterform}) and is tending to 1 from below at the bottom figure (case d) in case of the reaction \ce{2X + Y <=>[$k_1$][$k_{-1}$] 2Z}. \ce{Y} is in excess in both cases. \(V~=~1~\Lit,\) and other data are given in the text.}
\label{fig:rm22}
\end{figure}
\end{example}

Now we formulate our experience collected on several models. 
Consider reaction \eqref{eq:singledb}.
\begin{conjecture}\label{conj:iniratio}
The sign of the difference \(k_{-1}(\cb^0)^{\beta}-k_1(\cb^0)^{\alpha}\)
and the sign of \(\lim_{t\to0}\frac{\xi_1(t)}{\xi_{-1}(t)}\)
is the same.
\end{conjecture}

Convergence has been proved above.
The ratio at \(t=0\) is not defined, but  the limit of the ratio when \(t \to +0\)
can be calculated using the l'Hospital Rule as
\begin{equation*}
\lim_{t\to+0}\frac{\xi_{1}(t)}{\xi_{-1}(t)}=
\lim_{t\to+0}\frac{\dot{\xi}_{1}(t)}{\dot{\xi}_{-1}(t)}=
\frac{k_1}{k_{-1}}(\cb^0)^{\alphab-\betab},
\end{equation*}
and \(\frac{k_1}{k_{-1}}(\cb^0)^{\alphab-\betab}<1\) is equivalent to saying that 
\(k_1(\cb^0)^{\alphab}<k_{-1}(\cb^0)^{\betab}.\)

The meaning of the above conjecture is quite obvious: 
if the forward reaction proceeds slower at the beginning than the backward reaction,
then limit 1 of the ratio is approached from below etc.
The concept of stoichiometric initial condition seems to play no role here.

Instead of studying other simple reactions, we generalize the above result.
\subsubsection{Multiplication of the ratios.}
\begin{theorem}\label{thm:product}
If the complex chemical reaction \eqref{eq:revccr} is detailed balanced, i.e. \eqref{eq:dbgeneralcond} is fulfilled, then
\fbox{\(
\lim_{t\to +\infty}
\prod_{r=1}^R
\frac{\xi_{r}(t)}{\xi_{-r}(t)}=1.
\)}
\end{theorem}
\Pf{
It is similar to that of Proposition \ref{prop:specdb}: 
all the separate factors tend to 1 as \(t\to+\infty.\)}

\begin{example}
Consider the example in Fig. \ref{fig:rm3} that is not unconditionally detailed balanced.
\begin{figure}
\centering
\includegraphics[height=0.5\textheight]{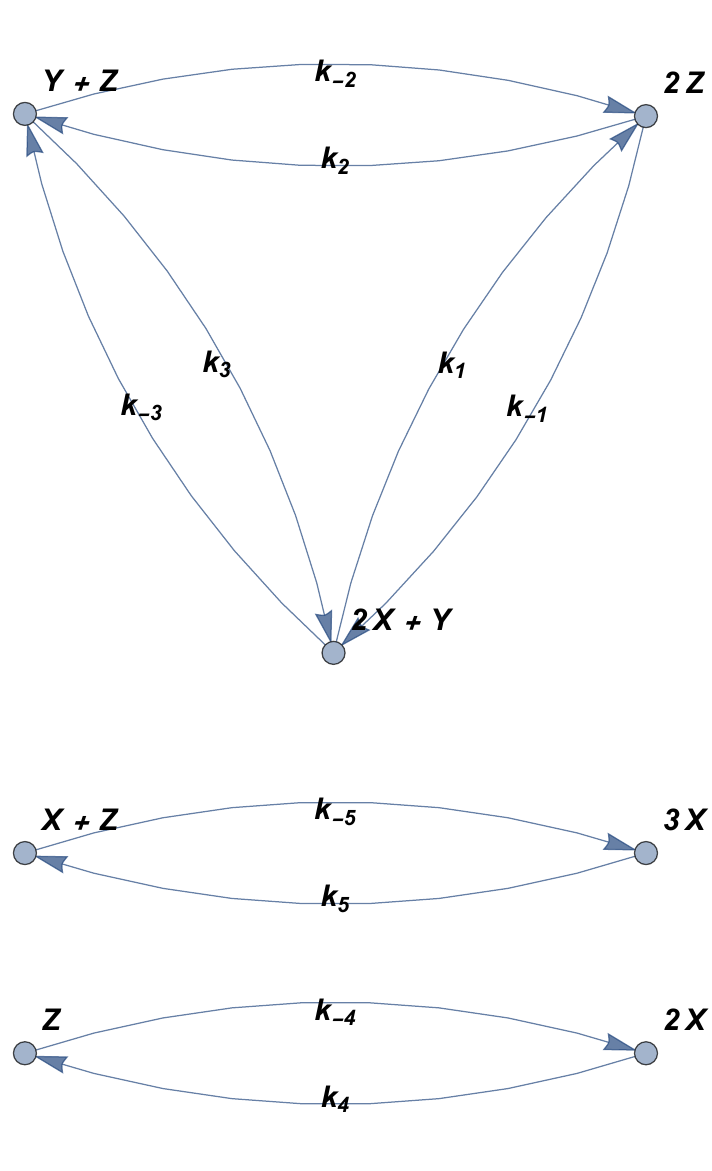}
\caption{A triangle coupled with a Wegscheider-type reaction\cite{wegscheider}}
\label{fig:rm3}\end{figure}
In this case, the condition of detailed balancing to hold is
\[
k_{-1} k_{-2} k_{-3}=k_1 k_2 k_3,\quad k_3 k_5=k_{-3} k_{-5},\quad k_3 k_4=k_{-3} k_{-4}
\]
as applying either the \emp{circuit conditions} and the \emp{spanning forest conditions} \cite{feinbergdb} or use the algebraic condition coming from the \emp{Fredholm alternative theorem} (see p. 133 of\cite{tothnagypapp}) gives. 
The calculations leading to similar equalities in more complicated cases
can be carried out using the function \mma{DetailedBalanced} of the package \mma{ReactionKinetics}, supplement to the book by T\'oth et al.\cite{tothnagypapp}. See an example on p. 136 of the cited book.
\begin{figure}
\centering
\includegraphics[width=0.7\linewidth]{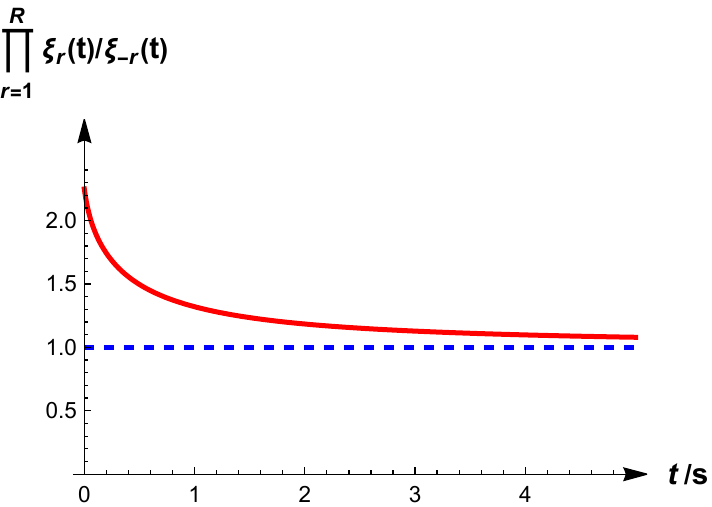}
\caption{Convergence of the product of ratios from above in case of reaction in Fig. \ref{fig:rm3} with parameters implying \emp{detailed balancing} as follows:
\(k_1~=~1~\Litsix~\mol^{-2}~\mathrm{s}^{-1}\), 
\(k_{-1}~=~2~\Lit~\mol^{-1}~\mathrm{s}^{-1}\), 
\(k_2~=~3~\Lit~\mol^{-1}~\mathrm{s}^{-1}\), 
\(k_{-2}~=~1~\Lit~\mol^{-1}~\mathrm{s}^{-1}\), 
\(\mathbf{k_3}~=~\frac{1}{3}~ \Lit~\mol^{-1}~\mathrm{s}^{-1}\),
\(\mathbf{k_{-3}}~=~\frac{1}{2}~ \Litsix~\mol^{-2}~\mathrm{s}^{-1}\),
\(k_4~=~3~\Lit~\mol^{-1}~\mathrm{s}^{-1}\),
\(k_{-4}~=~2~{\mathrm{s}}^{-1}\), 
\(k_5~=~3~\Litsix~\mol^{-2}~\mathrm{s}^{-1}\),
\(k_{-5}~=~2~\Lit~\mol^{-1}~\mathrm{s}^{-1}\), 
\(c_{\ce{X}}^0~=~c_{\ce{Y}}^0~=~c_{\ce{Z}}^0~=~1~\mL\), 
\(V~=~1~\Lit.\)}
\label{fig:rm30}
\end{figure}
Thus, Fig. \ref{fig:rm30} shows that the ratio tends to 1. 
\end{example}
\begin{example}
Even if the reversible reaction is \emp{not detailed balanced}, it will not blow up\cite{boroshofbauer}. 
The stationary point exists and is unique, and the product of the ratios will converge; although the limit will be different from 1, see Fig. \ref{fig:rm30notdb}. 
Let us use l'Hospital's Rule to show this, in the case of a single factor of the product:
\begin{equation*}
\lim_{t\to+\infty}\frac{\xi_r(t)}{\xi_{-r}(t)}=    
\lim_{t\to+\infty}\frac{\dot{\xi}_r(t)}{\dot{\xi}_{-r}(t)}=
\frac{\dot{\xi}_r(+\infty)}{\dot{\xi}_{-r}(+\infty)}=
\frac{k_r}{k_{-r}}\frac{(\cb^*)^{\alphab(.,r)}}{(\cb^*)^{\betab(.,r)}},
\end{equation*}
which is not necessarily equal to 1.

\begin{figure}
\centering
\includegraphics[width=0.7\linewidth]{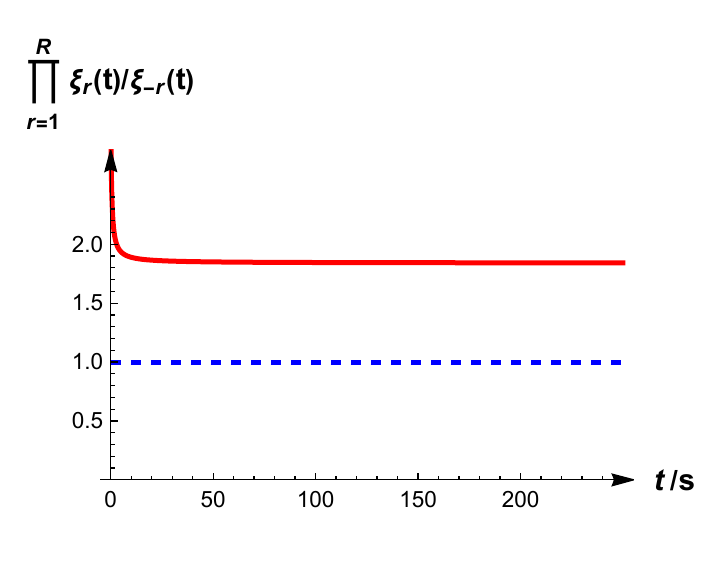}
\caption{Convergence of the product of ratios in the case of reaction in Fig. \ref{fig:rm3} with parameters \emp{not fulfilling detailed balance} as follows:
\(
k_1~=~1~\Litsix~\mol^{-2}~\mathrm{s}^{-1}\), 
\(k_{-1}~=~2~\Lit~\mol^{-1}~\mathrm{s}^{-1}\), 
\(k_2~=~3~\Lit~\mol^{-1}~\mathrm{s}^{-1}\), 
\(k_{-2}~=~1~\Lit~\mol^{-1}~\mathrm{s}^{-1}\),
\(\boldsymbol{k_3}~=~1~\Lit~\mol^{-1}~\mathrm{s}^{-1}\), 
\(\boldsymbol{k_{-3}}~=~1~\Litsix~\mol^{-2}~\mathrm{s}^{-1}\), 
\(k_4~=~3~\Lit~\mol^{-1}~\mathrm{s}^{-1}\),
\(k_{-4}~=~2~{\mathrm{s}}^{-1}\), 
\(k_5~=~3~\Litsix~\mol^{-2}~\mathrm{s}^{-1}\),
\(k_{-5}~=~2~\Lit~\mol^{-1}~\mathrm{s}^{-1}\), 
\(c_{\ce{X}}^0~=~c_{\ce{Y}}^0~=~c_{\ce{Z}}^0~=~1~\mL,\) 
\(V~=~1~\Lit.\)}
\label{fig:rm30notdb}
\end{figure}
\end{example}
\subsection{Reactions with an attractive stationary point}
All our observations can be summarized in the trivial proposition below. Before stating it, we need a definition for general ordinary differential equations.

Let \(M\in\N,\)  \(\fb\in\CC^1(\R^M,\R^M)\) and consider the initial value problem
\begin{equation}\label{eq:ivp}
\dot{\xb}(t)=\fb(\xb(t)), \xb(0)=\xb^0\, (\in\R^M).
\end{equation}
\begin{definition}
The stationary point \(\xb^*\) of \eqref{eq:ivp} is said to be \emp{attractive}, if all the solutions starting from a neighbourhood of \(\xb^*\) are defined for all positive time, and tend to it as \(t\to+\infty\).
\end{definition}
Note that attractiveness is less then being asymptotically stable and different from being stable. The neighbourhood mentioned in the definition is the \emp{domain of attraction} of the point \(\xb^*.\)

As a trivial reformulation of the definitions, we arrive at our general statement.
\begin{proposition}\label{prop:main}
Let \(\xb^*\) be an attractive stationary point of the initial value problem Eq. \eqref{eq:ivp}, and assume that \(\xb^0\) is located in the attracting domain of \(\xb^*\).
Assume furthermore, that for some
\(g\in\CC(\R^M,\R):g(\xb^*)\neq0,\) then \(\lim_{t\to+\infty}\frac{g(\xb(t))}{g(\xb^*)}=1.\)
\end{proposition}
This proposition is not useful enough. 
The reason is that in most cases the derivatives of the reaction extents are positive,
therefore, the reaction extents themselves tend to infinity, that is they have no (finite) stationary values. 
They can only have it when the derivative tends to zero like in the consecutive Example \ref{ex:convexconcave} above: we have seen that \(\xi_1(t),\xi_2(t)\to Vc^0_{\ce{X}}.\)

Now the problem arises that one has to find cases when some of the reaction rates tend to zero,
as this is a necessary (although not necessarily sufficient!) condition to have a finite reaction extent if \(t\to +\infty.\) 

Note that we did not use Proposition \ref{prop:main} when calculating the limit of the ratios of reaction extents.
\section{What if the conditions are not fulfilled?}
In the first part of the paper, we calculated the reaction extents for the reaction steps of simple reactions, for detailed balanced reactions, for reactions with a kinetic differential equation having an attracting stationary point, etc. 
Our main question in the present part is:
What happens if one takes an exotic reaction that has multiple stationary point(s), and shows oscillations or even chaos?
\subsection{Multistationarity}
\begin{figure}
\centering
\includegraphics[width=0.85\linewidth]{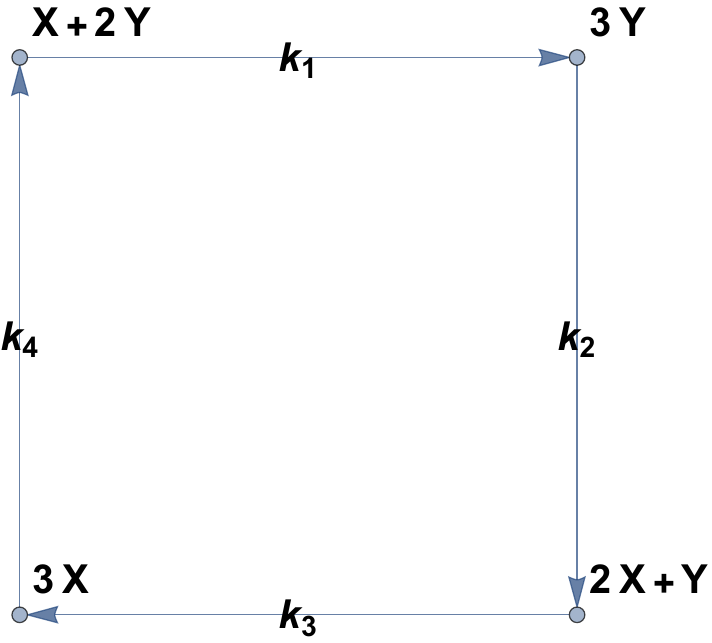}
\caption{The Horn--Jackson reaction network
with \(k_1=k_3=1~\Litsix~\mol^{-2}~\mathrm{s}^{-1}\) and \(k_2=k_4=k\). 
The value of \(k\) is varied as described in the text, 
its unit is the same as that of the other rate coefficients.}
\label{fig:hj0}
\end{figure}
\begin{figure}
\centering
\includegraphics[width=0.95\linewidth]{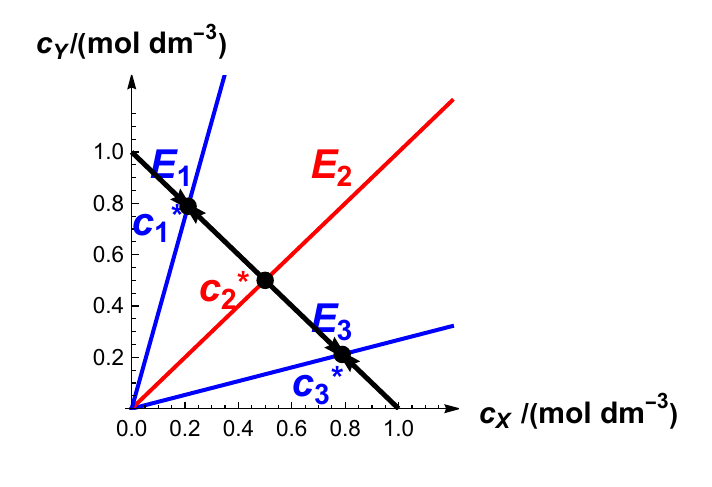}
\caption{Phase plane of the reaction network in Fig. \ref{fig:hj0}.
The black line is one of the compatibility classes, where the total concentration is unity. \(E_1, E_2\), and \(E_3\) are the sets of stationary points, and their intersections with the line of the given compatibility class results in the actual stationary points \(\cb^*_1, \cb^*_2\), and \(\cb^*_3\). Blue denotes asymptotic stability, red denotes instability. The black arrows show the direction of the motion along the trajectories in the neighborhood of the attracting stationary points.}
\label{fig:hj1}
\end{figure}

Horn and Jackson\cite{hornjackson}, (see p. 110) has shown that the complex chemical reaction in Fig. \ref{fig:hj0} has three (positive) stationary points in every stoichiometric compatibility class if 
the numerical value of \(k\) lies between 0 and \(\frac{1}{6}.\)
To be more specific, let us choose 
\(
k~=~\frac{1}{10}~\Litsix~\mol^{-2}~\mathrm{s}^{-1}\), and  \(c^0_{\ce{X}}~=~c^0_{\ce{Y}}~=~\frac{1}{2}~\mL.\) 
Then, easy calculation--neglecting the units for simplicity--shows that in the \emp{stoichiometric compatibility} class
\(\{[c_{\ce{X}}\quad c_{\ce{Y}}]; c_{\ce{X}}+c_{\ce{Y}}=1\}\)
(or, for cases when the total concentration is unity) there are three stationary points:
\begin{enumerate}
\item 
the stationary point
\(
\cb^*_1:=\begin{bmatrix}
\frac{1}{3 + \sqrt{3}}&\frac{3 + \sqrt{3}}{6}
\end{bmatrix}
\) is  globally asymptotically stable (i.e. attracting)  with the attracting domain 
\(\{[c_{\ce{X}}\quad c_{\ce{Y}}]: 0\le c_{\ce{X}}<\frac{1}{3 + \sqrt{3}}, c_{\ce{X}}+c_{\ce{Y}}=1\},
\)  and
\item 
the stationary point \(\cb^*_2:=[\frac{1}{2}\quad\frac{1}{2}]\)
is unstable (i.e. non-attracting),  and
\item 
the stationary point
\(
\cb^*_3:=\begin{bmatrix}
\frac{1}{3 - \sqrt{3}}&\frac{3 - \sqrt{3}}{6}
\end{bmatrix}
\) 
is globally asymptotically stable (i.e. attracting)
 with the attracting domain 
\(
\{[c_{\ce{X}}\quad c_{\ce{Y}}];\frac{1}{3 - \sqrt{3}}<c_{\ce{X}}\le1, c_{\ce{X}}+c_{\ce{Y}}=1\}
\).
\end{enumerate}
The character of the stationary points can be determined by using standard linear stability analysis.
Fig. \ref{fig:hj1} summarizes the behavior of the trajectories in the neighborhood of the stationary points. 

\begin{figure}
\centering
\includegraphics[width=0.85\linewidth]{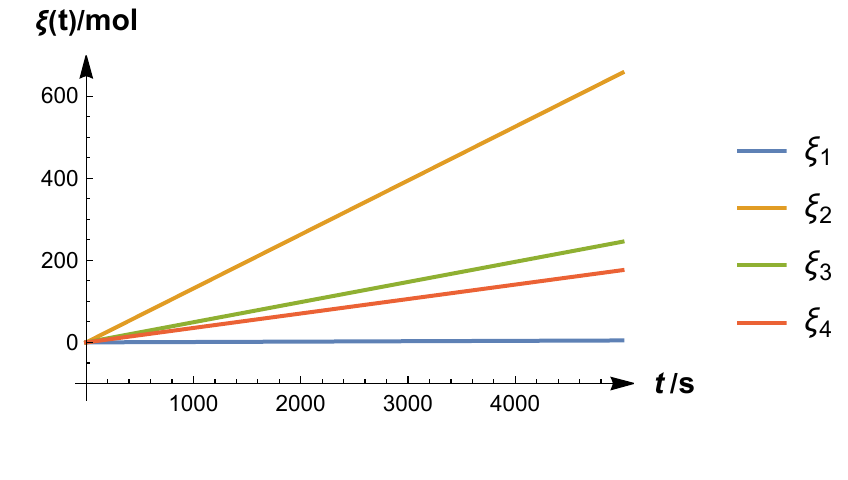}\\
\includegraphics[width=0.85\linewidth]{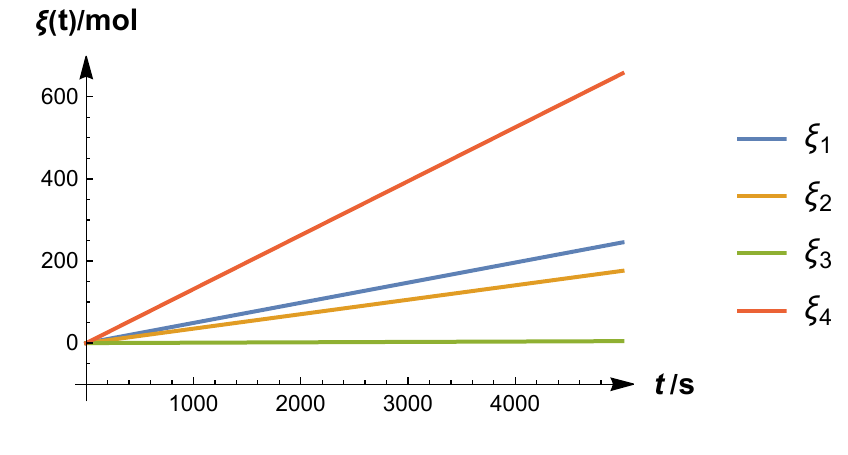}
\caption{Reaction extents starting from the attracting domain of \(\cb^*_1\) (above) and \(\cb^*_3\) (below). As there are large differences in the values,
the smallest one seems to be zero, but it is not.
The reaction steps in  Fig. \ref{fig:hj0} are numbered clockwise, starting with the reaction \ce{X + 2Y -> 3Y}. 
Note that the ranking of the reaction extents is different in the two cases.}
\label{fig:hj}
\end{figure}

The reaction extents tend to infinity in all cases,
but they are ordered differently for different initial conditions, see in Fig. \ref{fig:hj}.
This reflects the fact that not the same reactions are the fastest or slowest in the two cases.
\subsection{Oscillation}
We shall study here two oscillatory reactions. 
First, the often used Lotka--Volterra reaction\cite{lotka,volterra} comes that is not only theoretically interesting because it can be used to describe the oscillations in cold flames\cite{frankkamenetskii} or see Ref.\cite{frankkamenetskiiPUP}.
The experimentally based R\'abai reaction\cite{rabai} aimed at describing pH oscillations follows as the second. 
One may say that the Brusselator model\cite{prigoginelefever} would be a more realistic choice, as it results in  limit cycle solutions. However, it has a third-order step that makes the calculations more tedious.
The type of calculations shown below would give almost the same kind of results with the Brusselator, too.
\subsubsection{The Lotka--Volterra reaction.}
The irreversible and reversible cases behave in qualitatively different ways.

\paragraph{Irreversible case:}

\begin{figure}[!ht]
\centering
\includegraphics[width=0.85\linewidth]{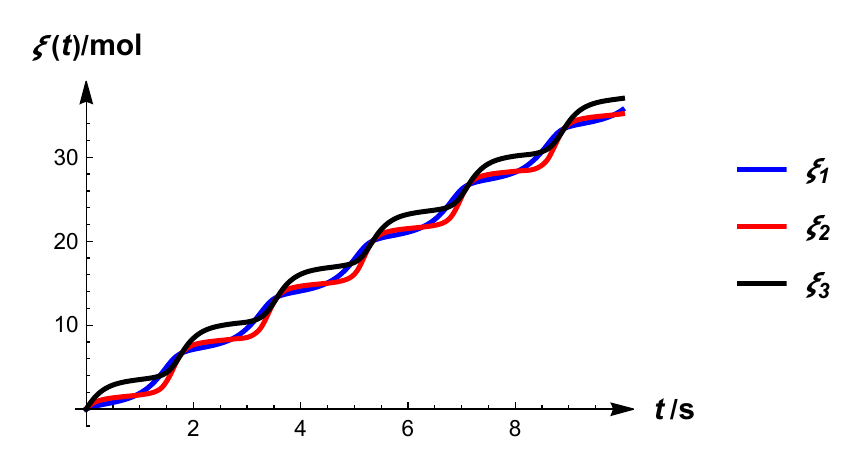}\\
\includegraphics[width=0.85\linewidth]{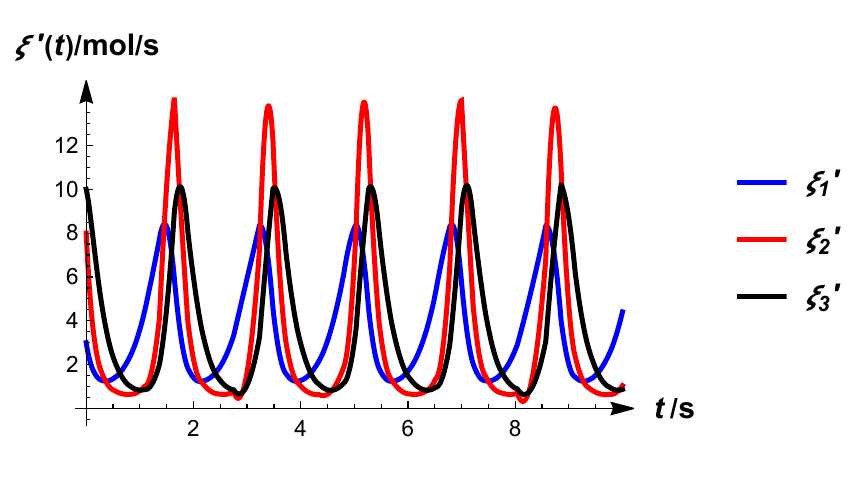}\\
\includegraphics[width=0.85\linewidth]{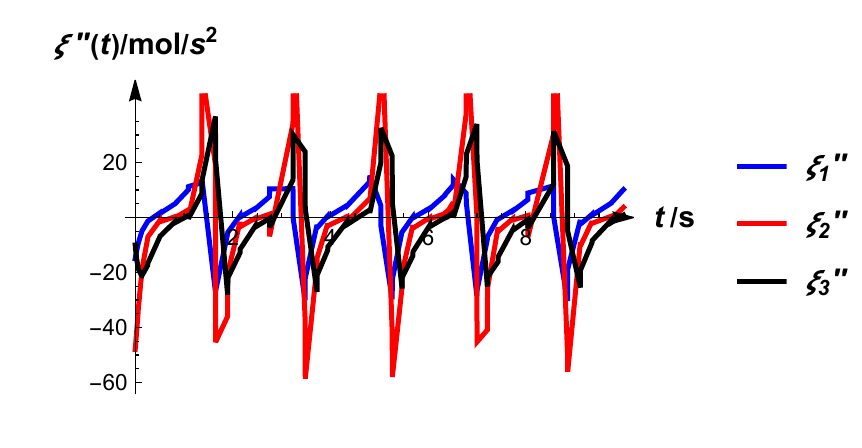}
\caption{The individual reaction extents and their first and second derivatives in case of the irreversible Lotka--Volterra reaction with 
\(k_1~=~3~{\mathrm{s}}^{-1}\),
\(k_2~=~4~\Lit~\mol^{-1}~\mathrm{s}^{-1}\), 
\(k_3~=~5~{\mathrm{s}}^{-1}\),
\(c^0_{\ce{X}}~=~1~\mL, c^0_{\ce{Y}}~=~2~\mL\), 
\(V~=~1~\Lit.\)}
\label{fig:pulsating}
\end{figure}

\begin{figure}[ht]
\centering
\includegraphics[width=0.75\linewidth]{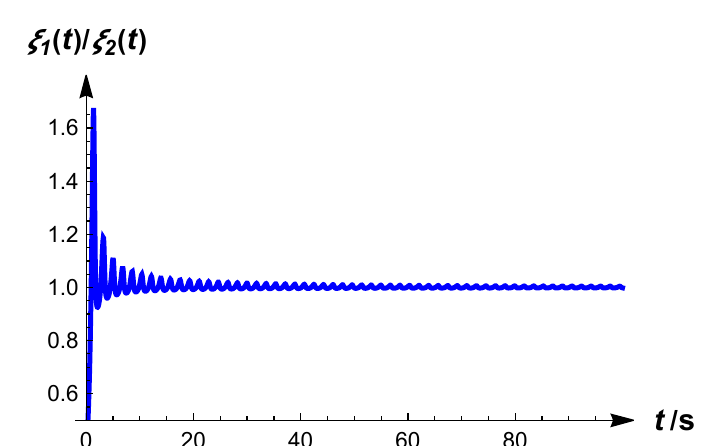}\\
\includegraphics[width=0.75\linewidth]{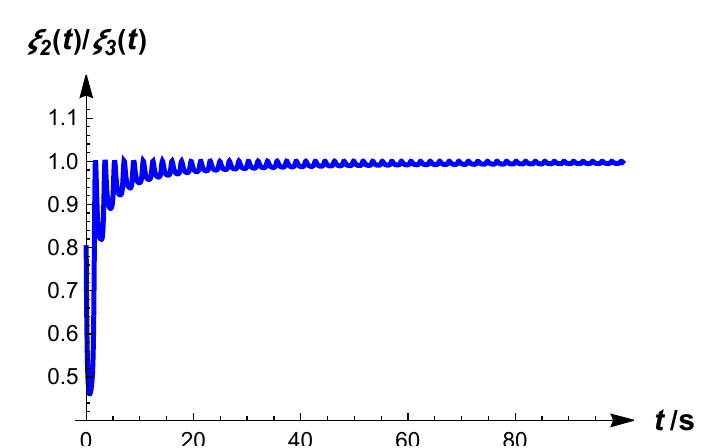}\\
\includegraphics[width=0.75\linewidth]{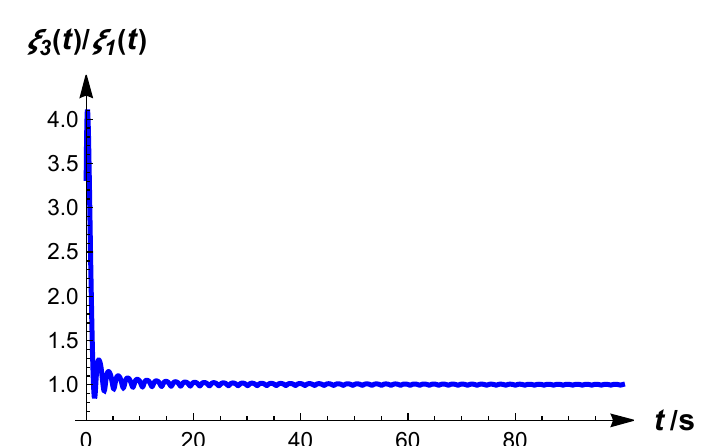}
\caption{The ratios of the reaction extents in case of the irreversible Lotka--Volterra reaction with the parameters as in Fig. \ref{fig:pulsating}.}
\label{fig:lotkaratios}
\end{figure}

It is known\cite{potalotka,schumantoth} that under some mild conditions the only two-species reaction to show oscillations is the (irreversible) Lotka--Volterra reaction \ce{X ->[$k_1$] 2X}, \ce{X + Y ->[$k_2$] 2Y}, \ce{Y ->[$k_3$] 0}. (Cf. also the paper by T\'oth and H\'ars\cite{tothhars} and that by Banaji and Boros\cite{banajiboros}.) 
It has a single positive stationary point that is stable but not attractive, therefore, one cannot apply Proposition \ref{prop:main} above.
Note that the individual reaction extents are not oscillating; they are "pulsating" while monotonously increasing to infinity. 
They have an oscillatory derivative, and the zeros of their second derivative clearly show the endpoints of the periods, 
see Fig. \ref{fig:pulsating}. 
It may be a good idea to calculate any kind of reaction extent \emp{for a period} in case of oscillatory reactions.
We are going to study this point later.
It is interesting to have a look at the ratios of the reaction extents, as they seem to tend 1, see Fig. \ref{fig:lotkaratios}.
We assume that this phenomenon is related to the fact that the oscillatory solution results in a closed curve in the phase plane of the irreversible Lotka--Volterra reaction.

\paragraph{Reversible case, detailed balanced:}

\begin{figure}[ht]
\centering
\includegraphics[width=0.85\linewidth]{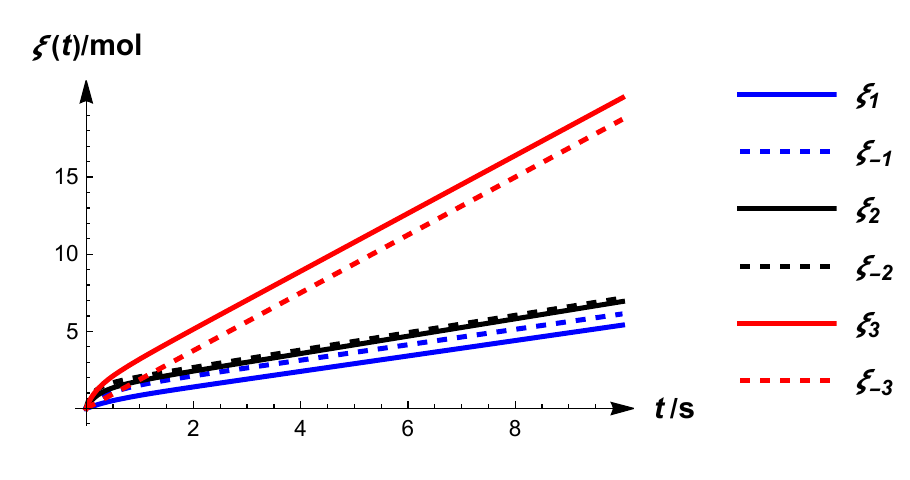}\\
\includegraphics[width=0.85\linewidth]{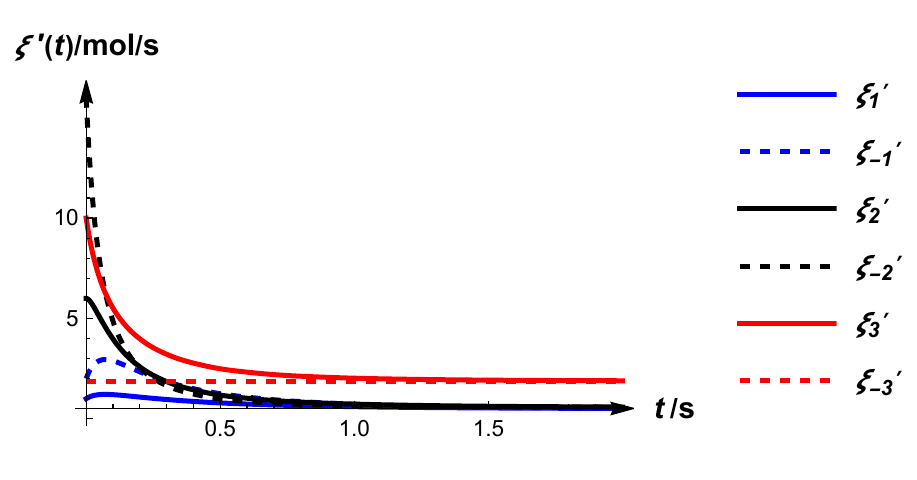}\\
\includegraphics[width=0.85\linewidth]{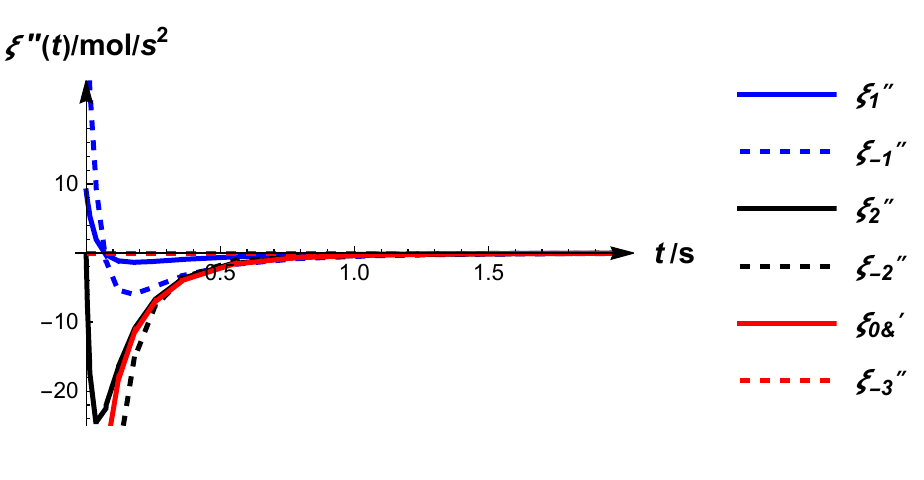}
\caption{The individual reaction extents and their first and second derivatives in case of the reversible, detailed balanced Lotka--Volterra reaction with 
\(k_1~=~1~{\mathrm{s}}^{-1}\),
\(k_{-1}~=~2~\Lit~\mol^{-1}~\mathrm{s}^{-1}\),
\(k_2~=~3~\Lit~\mol^{-1}~\mathrm{s}^{-1}\),
\(k_{-2}~=~4~\Lit~\mol^{-1}~\mathrm{s}^{-1}\),
\(k_3=~5~{\mathrm{s}}^{-1}\),
\(k_{-3}~=~\frac{15}{8}~\mol~\mathrm{dm}^{-3}~{\mathrm{s}}^{-1}\),
\(c^0_{\ce{X}}~=~1~\mL\), \(c^0_{\ce{Y}}~=~2~\mL\), \(V~=~1~\Lit.\)}
\label{fig:lvrevdb}
\end{figure}

The reversible Lotka--Volterra reaction
\ce{X <=>[$k_1$][$k_{-1}$] 2X}, 
\ce{X + Y <=>[$k_2$][$k_{-2}$] 2Y}, 
\ce{Y <=>[$k_3$][$k_{-3}$] 0}
is also worth studying.
First, let us note that for all values of the reaction rate coefficients it has a single, positive stationary point because the reaction steps are reversible. 
Therefore, the system is permanent\cite{simon,borosexistence}, i.e. the trajectories remain in a compact set.  
If the trajectories remain in a compact set, then they are either tending to a limit cycle, or the stationary point is asymptotically stable. 
The first possibility is excluded by the above-mentioned theorem by P\'ota\cite{potalotka}, thus it is only the second possibility that remains. 
Fig. \ref{fig:lvrevdb} shows the behavior of the individual reaction extents.
Let us note that both the existence and uniqueness of the stationary state also follow from the Deficiency One Theorem (see p. 106 in Feinberg\cite{feinbergbook}, or p. 176 in Tóth et al.\cite{tothnagypapp})

\begin{figure}[ht]
\centering
\includegraphics[width=0.85\linewidth]{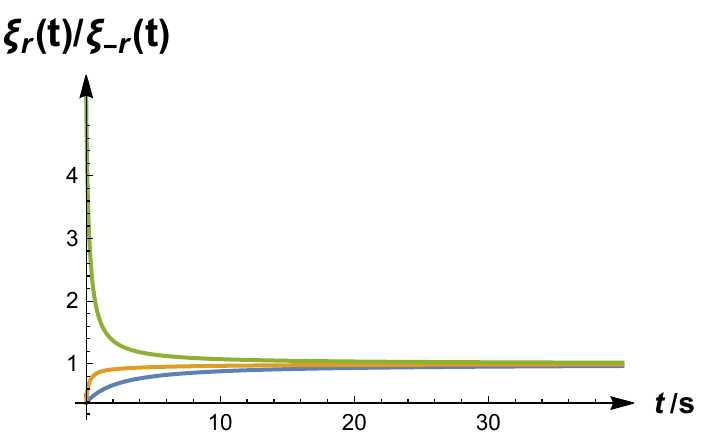}\\
\includegraphics[width=0.85\linewidth]{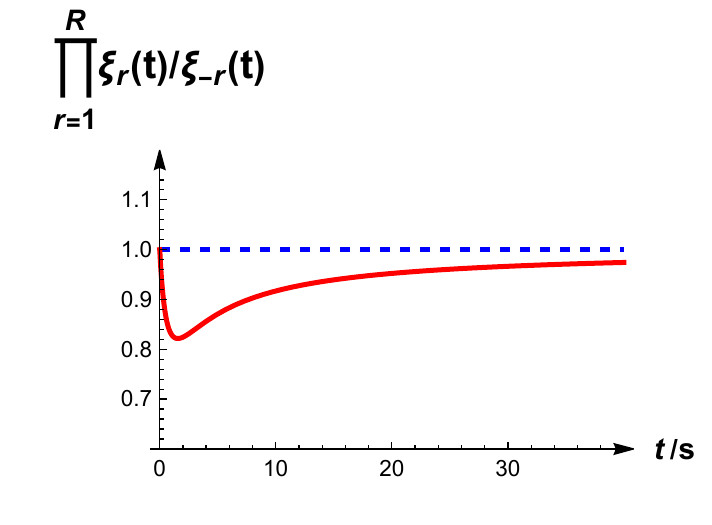}
\caption{
Above: Time evolution of the ratios of the individual reaction extents---blue for Reaction (1), orange for Reaction (2), and green for Reaction (3)---in case of the reversible, detailed balanced Lotka--Volterra reaction with
\(k_1~=~1~{\mathrm{s}}^{-1}\),
\(k_{-1}~=~2~\Lit~\mol^{-1}~\mathrm{s}^{-1}\), 
\(k_2~=~3~\Lit~\mol^{-1}~\mathrm{s}^{-1}\),
\(k_{-2}~=~4~\Lit~\mol^{-1}~\mathrm{s}^{-1}\),
\(k_3=~5~{\mathrm{s}}^{-1}\),
\(k_{-3}~=~\frac{15}{8}~\mol~\mathrm{dm}^{-3}~{\mathrm{s}}^{-1}\),
\(c^0_{\ce{X}}~=~1~\mL\), \(c^0_{\ce{Y}}~=~2~\mL\), \(V~=~1~\Lit.\)
Below: Time evolution of the product of the ratios tending to 1.}
\label{fig:lotkarevdbratios}
\end{figure}

If the reaction is detailed balanced that holds if and only if
\begin{equation}\label{eq:lvdb}
k_1k_2k_3=k_{-1}k_{-2}k_{-3}
\end{equation}
is true, then our Proposition 2 of the previous paper implies that
the product of the ratios of the reaction extents tends to 1, see Fig \ref{fig:lotkarevdbratios}. 
This follows also from our Theorem 3 there.

\paragraph{Reversible case, not detailed balanced:}

\begin{figure}[ht]
\centering
\includegraphics[width=0.85\linewidth]{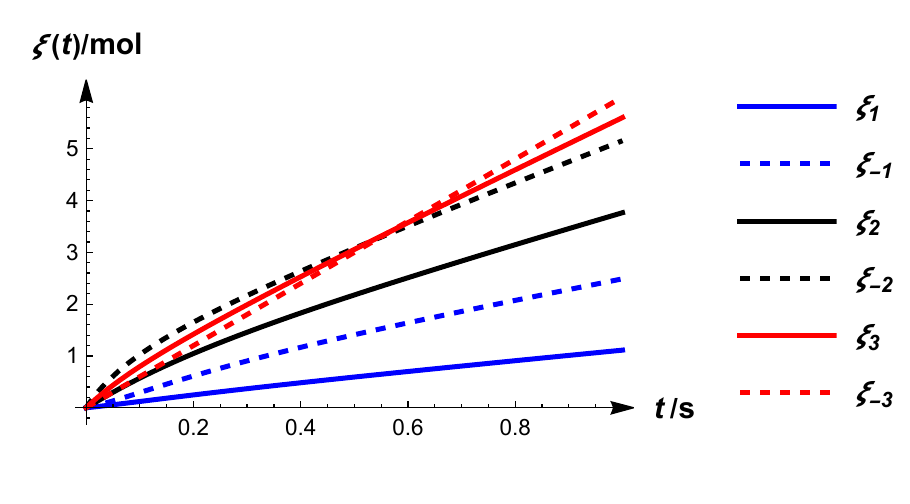}\\
\includegraphics[width=0.85\linewidth]{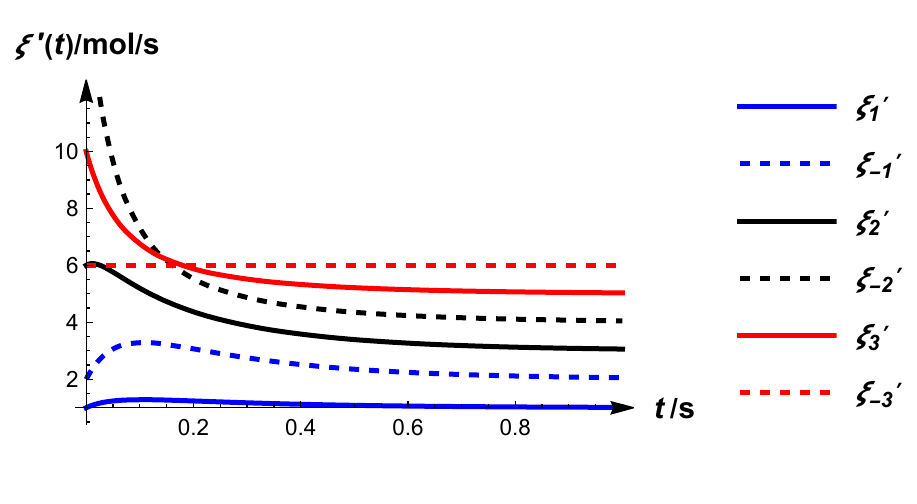}\\
\includegraphics[width=0.85\linewidth]{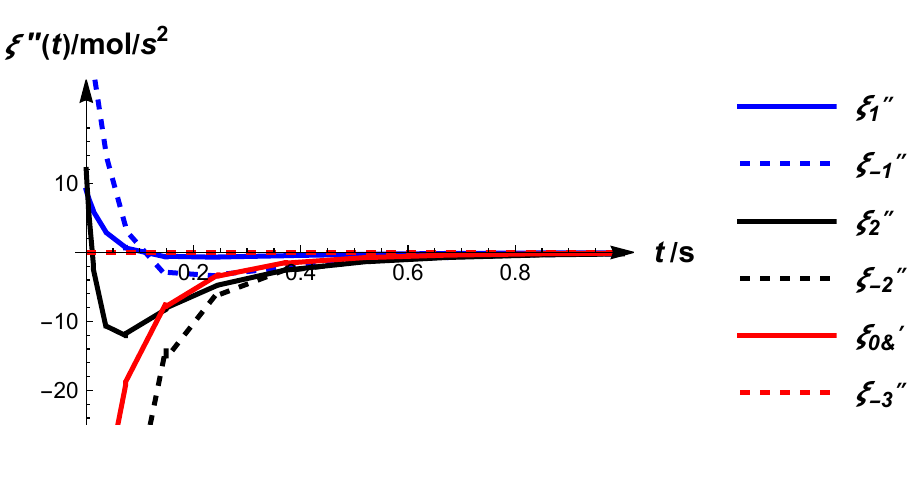}
\caption{The individual reaction extents and their first and second derivatives in case of the reversible, \emp{not} detailed balanced Lotka--Volterra reaction with \(k_1~=~1~{\mathrm{s}}^{-1},\)
\(k_{-1}~=~2~\Lit~\mol^{-1}~\mathrm{s}^{-1}\),
\(k_2~=~3~\Lit~\mol^{-1}~\mathrm{s}^{-1}\),
\(k_{-2}~=~4~\Lit~\mol^{-1}~\mathrm{s}^{-1}\),
\(k_3=~5~{\mathrm{s}}^{-1}\),
\(k_{-3}~=~6~\mol~\mathrm{dm}^{-3}~{\mathrm{s}}^{-1}\),
\(c^0_{\ce{X}}~=~1~\mL\), \(c^0_{\ce{Y}}~=~2~\mL\), \(V~=~1~\Lit.\)}
\label{fig:lotkanotdbxi}
\end{figure}

If Condition \eqref{eq:lvdb} does not hold, the reaction still has an \emp{attracting stationary point}. What is more, it has an asymptotically stable stationary point. 
Fig. \ref{fig:lotkanotdbxi} shows the behavior of the individual reaction extents.

\subsubsection{The R\'abai reaction of pH oscillation.}
Here we include a reaction proposed by R\'abai\cite{rabai} to describe pH oscillations. 
This reaction has much more direct contact with chemical kinetic experiments, and it is much more challenging---from the point of view of numerical mathematics---than the celebrated Lotka--Volterra reaction.

R\'abai\cite{rabai} starts with the steps
\begin{align*}
&\ce{A- + H+ <=>[$k_1$][$k_-1$] AH},\\
&\ce{AH + H+ +\{B\} ->[$k_2$] 2H+ + P-},
\end{align*}
where \ce{\{B\}} is an external species with a constant concentration. 
This reaction has a single stationary point
\begin{equation*}
c^*_{\ce{A-}}=0,\quad
c^*_{\ce{H+}}=c^0_{\ce{H+}}+c^0_{\ce{AH}},\quad
c^*_{\ce{AH}}=0,\quad
c^*_{\ce{P}}=c^0_{\ce{A-}}+c^0_{\ce{AH}}+c^0_{\ce{P}}
\end{equation*}
specializing into
\(
c^*_{\ce{A-}}=0,
c^*_{\ce{H+}}=c^0_{\ce{H+}},
c^*_{\ce{AH}}=0,
c^*_{\ce{P}}=c^0_{\ce{A-}}
\)
with the natural restriction on the initial condition
\(
c^0_{\ce{AH}}=0,
c^0_{\ce{P}}=0.
\)

\begin{figure}[ht]
\centering
\includegraphics[width=0.85\linewidth]{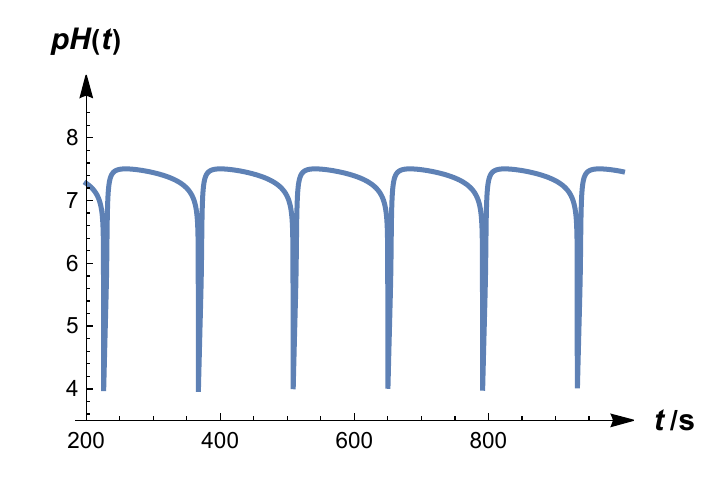}\\
\includegraphics[width=0.85\linewidth]{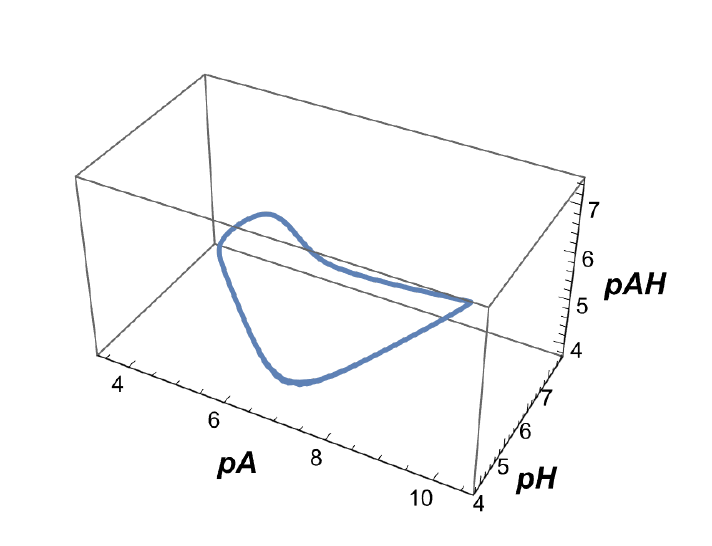}
\caption{Time evolution of the pH and the projection of the negative logarithm of the first three coordinates of the trajectory in case of the oscillating R\'abai reaction with 
\(k_1~=~10^{10}~\Lit~\mol^{-1}~\mathrm{s}^{-1}\),
\(k_{-1}~=~10^3~{\mathrm{s}}^{-1}\),
\(k_2~=~10^6~\Lit~\mol^{-1}~\mathrm{s}^{-1}\),
\(k_3~=~1~\Lit~\mol^{-1}~\mathrm{s}^{-1}\),
\(k_0~=~10^{-3}~{\mathrm{s}}^{-1}\),
\(c^0_{\ce{A-}}~=~5~\times~10^{-3}~\mL\),
\(c^0_{\ce{H+}}~=~10^{-3}~\mL\),
\(c^0_{\ce{AH}}~=~0~\mL\),
\(c^0_{\ce{P}}~=~0~\mL\),
\(V~=~1~\Lit.\)}
\label{fig:rabaiosc}
\end{figure}

Putting the reaction into a CSTR (continuously stirred flow-through tank reactor) 
means in the terms of formal reaction  kinetics that
all the species can flow out and some of the species may flow in, 
so that in the meantime the volume is maintained constant.
In the present case, the following steps are added
\begin{align}
&\ce{A- ->[$k_0$] 0},\label{rabaiout1}\\
&\ce{0 ->[$k_0c^0_{\mathrm{A}^-}$]A-},\label{rabaiin1}\\
&\ce{H+ ->[$k_0$] 0},\label{rabaiout2}\\
&\ce{0 ->[$k_0c^0_{\mathrm{H}^+}$] H+},\label{rabaiin2}\\
&\ce{AH ->[$k_0$] 0},\label{rabaiout3}
\end{align}
where $k_0$ is the volumetric flow rate normalized to the volume of the reactor (often called the reciprocal of the residence time) measured in unit \(\mathrm{s}^{-1}\). 
As a result of adding these steps, multistability may occur with appropriately chosen values of the parameters. When the reaction step
\begin{equation}\label{rabai3}
\ce{H+ + \{C$^-$\} ->[$k_3$] CH}
\end{equation}
is also added, one may obtain periodic solutions having appropriate parameter values, see Fig. \ref{fig:rabaiosc}. Let us remark that neither the R\'abai reaction nor the Lotka--Volterra reaction is mass-conserving.

It is instructive to cast a glance to the reaction extents in such a complex system.

\begin{figure}[ht]
\centering
\includegraphics[width=0.85\linewidth]{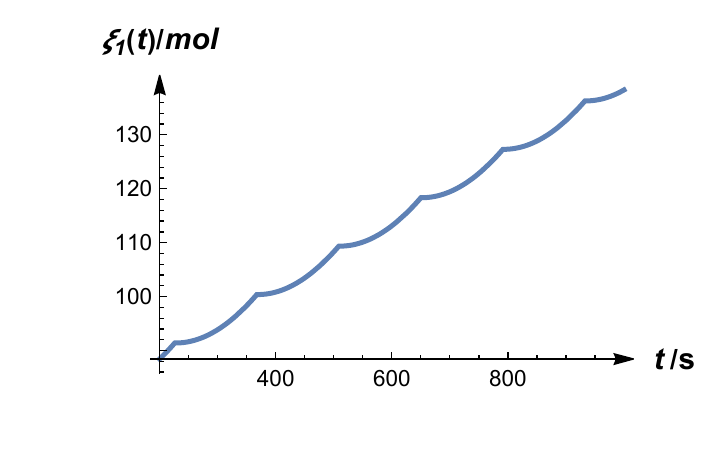}\\
\includegraphics[width=0.85\linewidth]{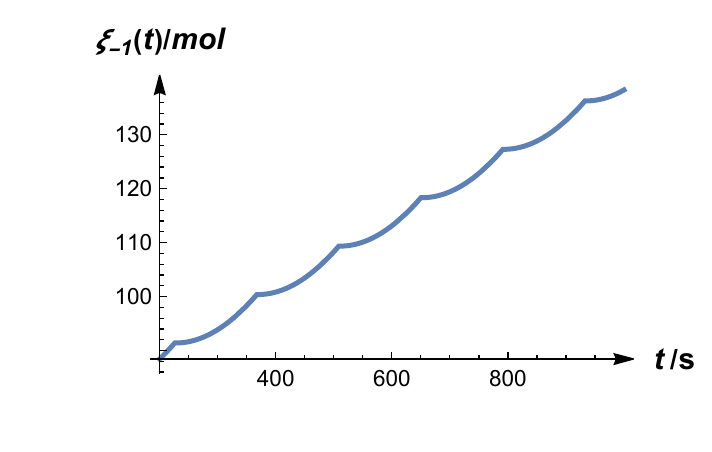}
\caption{Reaction extents of the forward (above) and backward (below) steps of the fast equilibrium reaction \ce{A- + H+ <=>[$k_1$][$k_-1$] AH} of the oscillating R\'abai reaction (in the same time window as that of in Fig. \ref{fig:rabaiosc}.)}
\label{fig:rabaioscext12} 
\end{figure}

\begin{figure}[ht]
\centering
\includegraphics[width=0.85\linewidth]{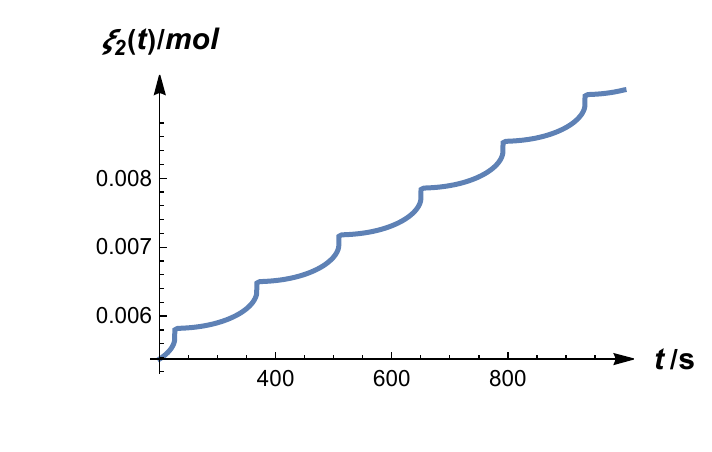}\\
\includegraphics[width=0.85\linewidth]{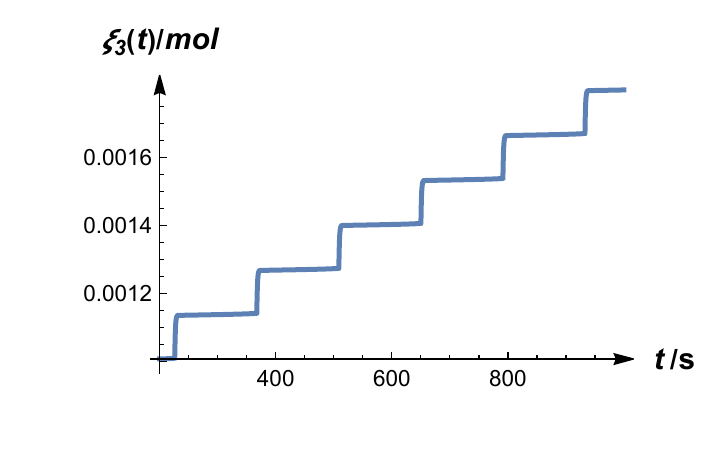}
\caption{Reaction extents of the reaction steps
\ce{AH + H+ +\{B\} ->[$k_2$] 2H+ + P-} (left) and 
\ce{H+ + \{C$^-$\} ->[$k_3$] CH} (right) of the oscillating R\'abai reaction (in the same time window as that of in Fig. \ref{fig:rabaiosc}.)}
\label{fig:rabaioscext23} 
\end{figure}
Note that the reaction extents of the fast equilibrium reaction \ce{A- + H+ <=>[$k_1$][$k_-1$] AH} shown in Fig. \ref{fig:rabaioscext12} are practically the same, or: their ratio tends to 1, as if no other steps were present! 
They are also four-five orders of magnitude higher than those of the auto-catalytic production \ce{AH + H+ +\{B\} ->[$k_2$] 2H+ + P-} and the slow pseudo first-order chemical removal \ce{H+ + \{C$^-$\} ->[$k_3$] CH} of \ce{H+} ion shown in Fig. \ref{fig:rabaioscext23}.
Note also the step-wise increase of reaction extent \(\xi_3(t)\).
\subsection{Chaos}
\begin{figure}[ht]
\centering
\includegraphics[width=0.85\linewidth]{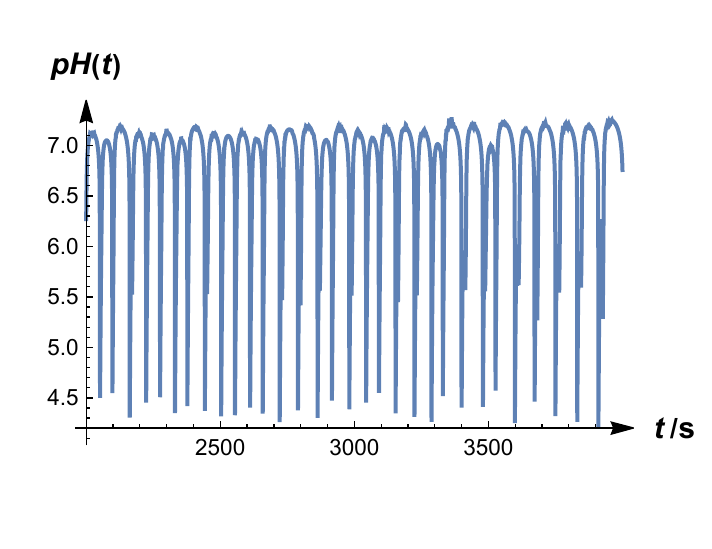}\\
\includegraphics[width=0.85\linewidth]{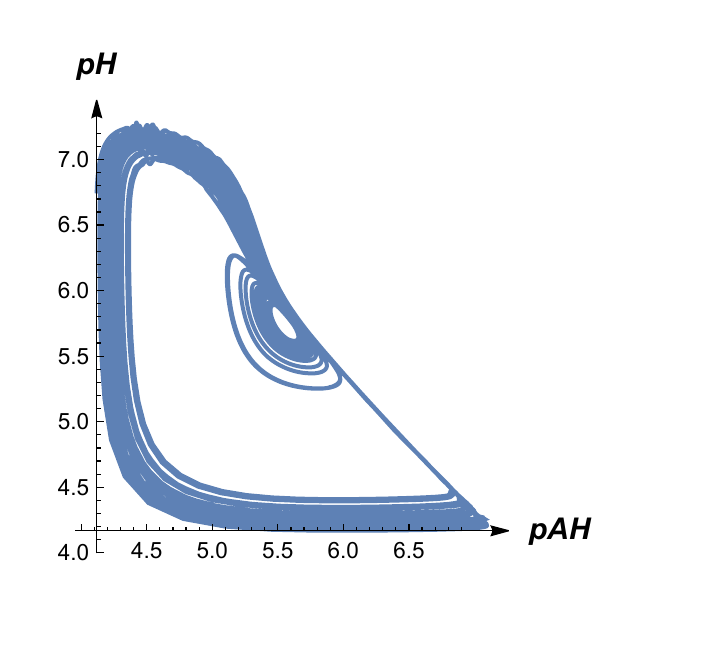}
\caption{Time evolution of the pH and the projection of the negative logarithm of the first two coordinates of the trajectory in case of the chaotically oscillating R\'abai reaction with 
the same parameters as in Fig. \ref{fig:rabaiosc} and
\(k_{-3}~=~1.5~\times~10^{-2}~{\mathrm{s}}^{-1}\),
\(k_{4}~=~~5~\times~10^{-2}~{\mathrm{s}}^{-1}.\) 
}
\label{fig:rabaichaos} 
\end{figure}

Here we use a version of the R\'abai reaction that \emp{can numerically be shown} to exhibit chaotic behavior, see Fig. \ref{fig:rabaichaos}. 
This is a good model for experimental pH oscillators also showing behavior that seems to be chaotic according to the usual standards.

\begin{figure}[ht]
\centering
\includegraphics[width=0.85\linewidth]{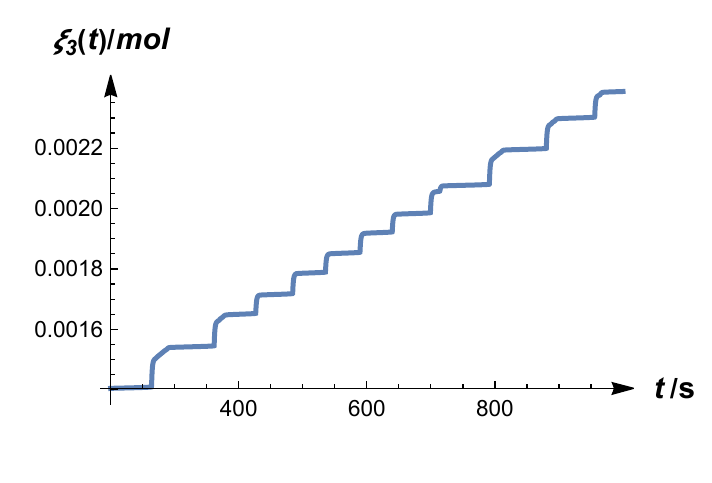}
\caption{Time evolution of the reaction extent 
of the forward step of reaction \eqref{rabai5}
in case of the chaotic R\'abai reaction with the same parameters as in Fig. \ref{fig:rabaichaos}.}
\label{fig:rabaichaosextent}
\end{figure}

When the reaction step \eqref{rabai3} is made reversible  
\begin{equation}\label{rabai5}
\ce{H+ + \{C$^-$\} <=>[$k_3$][$k_{-3}$] CH},
\end{equation}
and one also introduces both the chemical "removal" and the outflow of \ce{CH} 
\begin{equation}\label{rabai6}
\ce{CH ->[$k_{4}$] Q},
\end{equation}
\begin{equation}\label{rabai7}
\ce{CH ->[$k_{0}$] 0},
\end{equation}
chaotic solutions are obtained by using appropriate parameters and favorable input concentrations. Fig. \ref{fig:rabaichaos} illustrates this fact. The reaction extents tend to \(+\infty\) (see for example Fig. \ref{fig:rabaichaosextent}) in such a way that their derivative is chaotically oscillating (not shown), as expected.

\section{Conclusions}

A generalized definition for the reaction extent has been given by Bowen\cite{bowen} (included in Chapter 6 of the book edited and partially written by Truesdell\cite{truesdell}).
Another definition that turned out to be much better fitting into the framework of modern formal reaction kinetics in the last 50 years was given by Aris\cite{arisprolegomena1}. 
Still, neither of them became popular among chemists and chemical engineers. Our goal here is to further generalize the definition (essentially by Aris) to make it compatible with the present theory of reaction kinetics. The result will reveal that there existed a kind of sleeping definition with no use in chemical kinetics, and we show that this should not be the case.

We have introduced the concept of reaction extent for reaction networks of arbitrary complexity (any number of reaction steps and species) without assuming mass action kinetics. 
The newly defined reaction extent gives the advancement of each individual irreversible reaction step; 
in case of reversible reactions, we have a pair of reaction extents. 
In all the practically important cases, the  fact that the reaction extent is strictly monotonously increasing, implies that the reaction events never stop. 
This observation sheds new light onto the concept of \emp{dynamic equilibrium},
without alluding to either thermodynamics or statistical mechanics.

After a few statements on the qualitative behaviour of the reaction extent we made efforts to connect the notion with the traditional ones. 
Thus, we have shown that if the number of reaction steps is one, 
the reaction extent in the long run (as \(t\to+\infty\))
tends to 1 if appropriately scaled.
We have not used the expression \emp{progress of reaction}, 
and even less the \emp{reaction coordinate}. 
We agree that it is convenient to accept the proposal by Aris\cite{aris} to work with \(\frac{1}{V}\xib\), that is usually called the \emp{degree of advancement}. 
We have also shown that for an arbitrary number of reversible detailed balanced reaction steps the product of the ratios of the individual reaction extents also tends to 1 as \(t\to+\infty.\)

Our most general statement follows for arbitrary reactions having an attracting stationary point
and with a function not vanishing on the stationary point: in this case the value of the chosen function 
along the time dependent concentrations divided by the value of the given function at the stationary concentration tends to 1. 
Thus, this statement is true not only for the reaction extent but also for any appropriate functions.

One should take into consideration that although in the practically interesting cases, when the number of equations \(R\) in \eqref{eq:rmdiffegy} for the reaction extents is larger than \(M\), i.e. the number of the kinetic differential equations in \eqref{eq:ikdegen}: \(R > M\), then
the equations for the reaction extents are of a much simpler structure, 
as the right hand side of the differential equations \eqref{eq:rmdiffegy} describing them consist only of a single term. 
During calculations, we had the experience that it was numerically less demanding to solve the system of differential equations of the reaction extents than those of the concentrations. 

The main advantage of the new definition of reaction extent is that by knowing the kinetic model of a reacting system one can now calculate not only the time evolution of the concentration of each reacting species but also the number of occurrences of the individual reaction events.

As a by-product we have given an exact definition of the stoichiometric initial concentrations and the initial concentration in excess. 
One can  say that the concept of the newly defined reaction extent can be usefully applied to a larger class of reactions than usual, but in some (exotic) cases its use needs further investigations,
this we will start in the forthcoming paper.
It is for the reader to decide if we succeeded in avoiding all the traps mentioned in the Introduction. 
Quite a few authors treat the methodology of teaching the concept\cite{garst,moretti,vandezandevandergrienddekock};
we think this approach will only have its \textit{raison d'\^{e}tre} when the scientific background will have been clarified and agreed on.

Let us mention a few limitations and future directions of research. We have assumed throughout that volume (together with temperature and pressure) is constant. Tacitly, we assumed that we deal with homogeneous kinetics; heterogeneous systems are not taken into consideration. Also, we have not dealt with reaction-diffusion systems. 
We mention that recently, Peka\v{r}\cite{pekar} and Rodrigues et al.\cite{rodriguesbilleterbonvin} have applied the concept of reaction extent to the case when diffusion is also present. 
We have also mentioned a few mathematical conjectures that are to be investigated later. 

\section*{Supporting Information}
The file FiguresandCalculationGasparToth.pdf contains all the calculations and drawings made using the Wolfram language. 
Interested readers may request from the authors the .nb file usable for calculations.

\section*{Author Contributions}
The authors equally participated in all parts of the paper.

\section*{Conflicts of interest}
There are no conflicts to declare.

\section*{Acknowledgements}
The present work has been supported by the National Office for Research and Development  (2018-2.1.11-T\'ET-SI-2018-00007 and FK-134332). 
JT is grateful to Dr. J. Karsai (Bolyai Institute, Szeged University) and for Daniel Lichtblau (Wolfram Research) for their help. 
Members of the Working Committee for Reaction Kinetics and Photochemistry, 
especially Profs. T. Tur\'anyi and G. Lente, furthermore Drs. Gy. P\'ota and T. Nagy, made a number of useful critical remarks.

\section*{Notations}
Some of the readers may appreciate that we have collected the used notations.

\begin{table*}
\caption{Notations}
\label{tbl:notations1}
\begin{tabular*}{\textwidth}{@{\extracolsep{\fill}}clcc}
\hline
Notation & Meaning & Unit & Typical value \\
\hline\([a, b[\)&left-closed, right-open interval&&\\
\hline\(\Longrightarrow\)&implies&&\\
\hline\(\in\)&belongs to&&\\
\hline\(\forall\)&universal quantifier&&"for all"\\
\hline\(\exists\)&existential quantifier&&"there is"\\
\hline\(\odot\)&coordinate-wise product of vectors&&\\
\hline\(\Ab\T\)&transpose of the matrix \(\Ab\)&&\\
\hline\(c_m, c_{\ce{X}}\) & concentration of \ce{X($m$)} and \ce{X} & \(\mL\)& \\
\hline\(\cb\) &vector of concentrations  & & \\
\hline\(c_m^0\) & initial concentration of \ce{X($m$)} & \(\mL\)& \\
\hline\(c_m^*\) & stationary concentration of \ce{X($m$)} & \(\mL\)& \\
\hline\(\CC^i(A,B) \) & \(i\) times continuously differentiable \\
  & functions from \(A\) into \(B\) & & \\
\hline\(\Dom(u)\) & the domain of the function \(t\mapsto u(t)\)  & &\\
\hline \(J\subset\R\) & an open interval &&\\ 
\hline \(k, k_r,k_{-r}\) & reaction rate coefficient &\((\mL)^{1-\sum_{m=1}^{M}\alpha_{m,r}}~{\mathrm{s}}^{-1}\)&\\
\hline \(k_{0}\) & normalized volumetric flow rate& \({\mathrm{s}}^{-1}\) &\\
\hline\(M\) & the number of chemical species & & \(\in\N\)\\
\hline\(n_m\) & the quantity of species \ce{X($m$)}&\(\mol\) & \(\in\N\)\\
\hline\(n_m^0\) & the initial quantity of species \ce{X($m$)}  &\(\mol\) & \(\in\N\)\\
\hline\(\nb\) & the vector of the quantity of species & &\\
\hline\(\nb^0\) & the vector of the initial  quantity of species\\
\hline \(\N\) & the set of positive integers &&\\
\hline \(\N_0\) & the set of non-negative integers &&\\
\hline\(rate_r\) & rate of the \(\rth\) reaction step &\(\mL~\mathrm{s}^{-1}\) & \\
\hline\(\rateb\) & vector of reaction rates & & \\
\hline \(\R\) & the set of  real numbers&&\\
\hline \(\R^+\) & the set of positive real numbers &&\\
\hline \(\R^+_0\) & the set of non-negative real numbers &&\\
\hline\(t\) & time & s &  \(\in\R\)\\
\hline \(V\) & volume & \(\Lit\) & \\
\hline \(W_r\) & number of occurrences of the \(\rth\) step & &\(\in\N_0\) \\
\hline \(\Wb\) & vector of number of occurrences &&\\
\hline \(x_m\) & \(\mth\) dependent variable in a &  &\\
& differential equation&&\\
\hline\(\xb\) & vector of variables in a &  &\\
& differential equation &&\\
\hline\(\ce{X},\ce{Y},\ce{X($m$)}\) &chemical species  & & \\
\hline\(\alpha_{m,r},\alpha_{m}\) & stoichiometric coefficient & 1 & \(0,1,2,3\) \\
& in the reactant complex &  & \\
\hline \(\alphab\)  & matrix of reactant complex vectors  & & \\
\hline \(\beta_{m,r},\beta_{m}\) & stoichiometric coefficient & 1 & \(0,1,2,3\) \\
& in the product complex &  & \\
\hline \(\betab\)  & matrix of  product complex vectors & & \\
\hline \(\gamma_{m,r},\gamma_{m}\) &  stoichiometric number & 1 & \(-3,-2,\dots,2,3\) \\
\hline \(\gammab\) & stoichiometric matrix &  &\\
\hline \(\xi_r\) & reaction extent of the \(\rth\) step &\(\mol\)&\\
\hline \(\xib\) &vector of reaction extents & & \\
\hline
\end{tabular*}
\end{table*}




\renewcommand\refname{References}

\bibliography{_GasparToth}
\bibliographystyle{rsc} 
\eject
\section{Appendix}
\Pf{of Theorem \ref{thm:basics}.
\begin{enumerate}
\item 
This follows from Eq. \eqref{eq:specre}.
\item 
As  \(\cb\) is the solution of a differential equation with a continuously differentiable 
right-hand-side, it is twice continuously differentiable.
The definition \eqref{eq:specre} implies \(\xib\in\CC^2(J,\R^R).\)
\item
Take the derivative of Eq. \eqref{eq:specre} and use the third equation of \eqref{eq:trivicons}. 
The initial condition also comes from Eq. \eqref{eq:specre}.
\item
The derivative of \(\xi_r\) is non-negative. 
\item
If the derivative of \(\xi_r\) is positive for all \(t\in J\), then \(\xi_r\) is strictly monotonously increasing. 
If it is zero at some time \(t_0\in J,\)
then for some \(m: \alpha_{m,r}>0\) and \(c_m(t_0)=0.\)
However, \(c_m(t_0)=0\) can only hold if \(c_m(0)=0\)
held at the beginning, because an initially positive concentration cannot turn into zero, see
Theorem 1 on p. 617 in the book by Volpert and Hudyaev.\cite{volperthudyaev}
But then \(c_m(t)=0\) for all \(t\in J.\) 
Thus in this case for all \(t\in J : \dot{\xi}_r(t)=0,\) this, however, together with the initial condition \(\xi_r(0)=0\) implies that for all \(t\in J : \xi_r(t)=0\); therefore \(\xi_r\) is not strictly monotone, it is constant zero in this case. 
\end{enumerate}
}
\Pf{of Theorem \ref{thm:singlestep}.
\begin{enumerate}
\item 
The condition in Eq. \eqref{eq:cannot} implies that
\(\forall t\in\R^+_0\) and for all such \(m=1,2,\dots,M\) for which \(\alpha_m\neq0\) one has \(c_m(t)=0\), thus the right-hand-side of Eq. \eqref{eq:ximone} is zero. This, together with the initial condition \(\xi(0)=0\) proves the statement. 
\item 
The condition in Eq. \eqref{eq:produced} implies that
the right-hand-side of Eq. \eqref{eq:ximone} is positive,
and 
\begin{align*}
&\forall t\in\Dom(\xi):\\
&\ddot{\xi}(t)=k\sum_{m=1}^{M}\frac{\alpha_m\gamma_m}{c_m^0+\gamma_m\xi(t)/V}\prod_{p=1}^{M}(c_p^0+\gamma_p{\xi(t)}/{V})^{\alpha_p}>0,
\end{align*}
\begin{equation*}
\forall t\in\Dom(\xi):\ddot{\xi}(t)=\sum_{m=1}^{M}\frac{\alpha_m\gamma_m}{c_m^0+\gamma_m\xi(t)/V}\frac{\dot{\xi}(t)}{V} >0,
\end{equation*}
thus \(\xi\) is not only strictly monotonously increasing 
but also strictly convex from below proving the statement.
\item 
To calculate a limit when \(t\to+\infty\) we show that 
\begin{equation}\label{eq:supdom}
\sup(\Dom(\cb))=+\infty\text{ if and only if }\sup(\Dom(\xi))=+\infty.
\end{equation} 
If all the species are consumed then \eqref{eq:supdom} is obviously true.
Suppose that not all the species are consumed, e.g. suppose \ce{X($m$)} is consumed and \ce{X($p$)} is produced. Then the induced kinetic differential equation of the reaction Eq. \eqref{eq:onestep} implies that
\begin{equation*}
    \frac{\dot{c}_m}{-\gamma_m}+\frac{\dot{c}_p}{\gamma_p}=0,
\end{equation*}
therefore
\begin{equation*}
    -\frac{{c}_m(t)}{\gamma_m}+\frac{{c}_p(t)}{\gamma_p}=-\frac{{c}_m(0)}{\gamma_m}+\frac{{c}_p(0)}{\gamma_p}>0,
\end{equation*}
and this, together with the positivity of the concentrations implies \eqref{eq:supdom}.

If \ce{X($m$)} is consumed, then \(\gamma_m<0,\) therefore the factor
\(\left(c_m^0+\gamma_m\frac{\xi}{V}\right)^{\alpha_m}\)
in the right-hand-side of Eq.\eqref{eq:ximone}
is zero for \(\xi=-\frac{Vc_m^0}{\gamma_m}.\)
Since the solution of Eq. \eqref{eq:ximone} starts from zero and it is strictly monotonous, therefore it should tend to the smallest zero locus of the right-hand side proving the third part of the statement.

\end{enumerate}}
\begin{figure}
\centering
\includegraphics[width=0.3\linewidth]{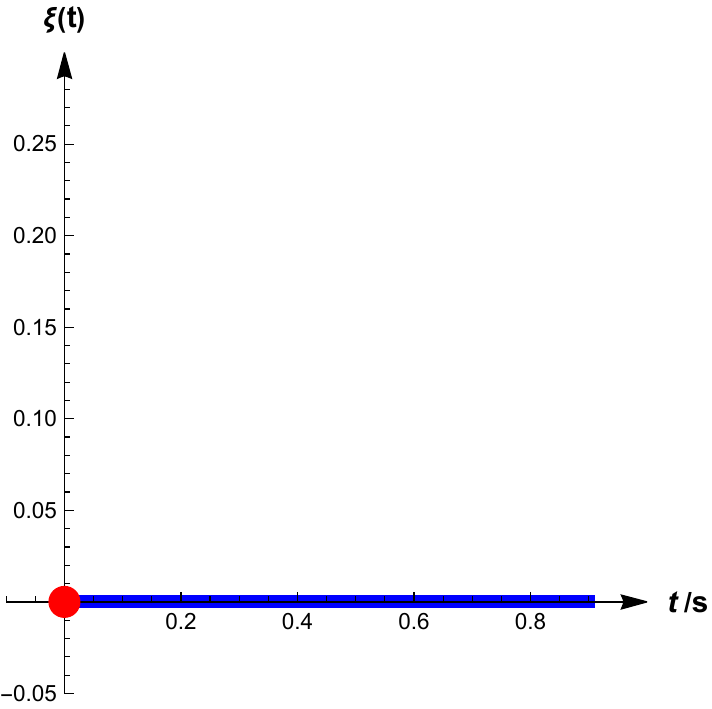}\quad
\includegraphics[width=0.3\linewidth]{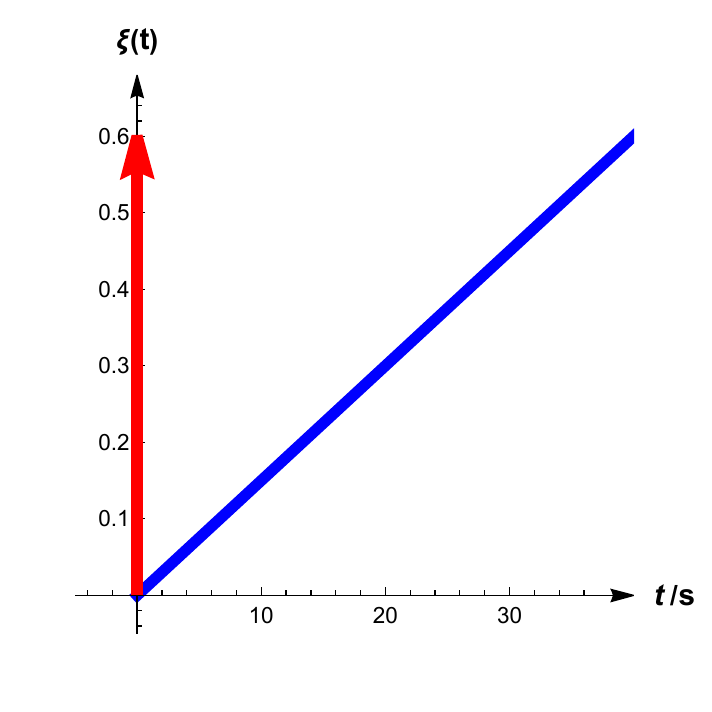}\quad
\includegraphics[width=0.3\linewidth]{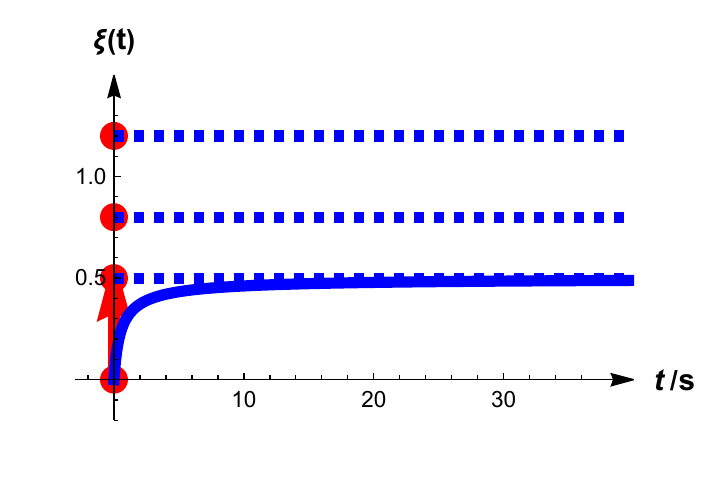}
\caption{The different possible structures of the phase line of Eq. \eqref{eq:ximone} are shown.
Left: zero is the single stationary point. Middle: there is no stationary point. Right: there are stationary points (at least one; here there are three). 
The stationary points and trajectories in the phase line representation (vertical axis) are red, while the solutions as a function of time (horizontal axes) are blue. Dashed blue in the right figure denotes the heights corresponding to the stationary points--two of them are never reached from the chosen initial value (0).}
\label{fig:schema}
\end{figure}
The proof can be considered as an application of phase-line analysis as visualized in Fig. \ref{fig:schema}.
The phase space of the equation \eqref{eq:ximone} is the half line of non-negative real numbers.
The following cases can occur: 
\begin{enumerate}
\item 
The origin is a stationary point, then the only solution is constant zero. 
It will turn out later that this case is less irrelevant from the chemical point of view when one has more than one reaction step.
\item 
There is no stationary point. 
Then the derivative of the reaction extent is always positive, thus it is strictly monotonously increasing.
Obviously, it cannot have a finite limit (it cannot stop), because if it would then the limit would be a stationary point, a contradiction.
\item 
There are stationary points (one or more). 
Then the solution to  \eqref{eq:ximone}---as it starts from the origin and is strictly monotonously increasing---will have as its limit the smallest stationary points as it has been said in the Theorem.

\end{enumerate}
\Pf{of Proposition \ref{prop:specdb}. 
First of all, let us note that the concentrations (and therefore the reaction extents) do not blow up\cite{boroshofbauer}, i.e. \(\sup(\Dom(\cb))=+\infty.\)
Next, as 
\(\lim_{t\to+\infty}{\xi}_1(t)=+\infty\)
and \(\lim_{t\to+\infty}{\xi}_{-1}(t)=+\infty\), 
and
\(\lim_{t\to+\infty}\dot{\xi}_1(t)=Vk_1(\cb^*)^{\alphab}\)
and \(\lim_{t\to+\infty}\dot{\xi}_{-1}(t)=Vk_{-1}(\cb^*)^{\betab}=Vk_1(\cb^*)^{\alphab}\), 
one can apply l'Hospital's Rule to get the desired result.}

\eject
\section*{Graphical Abstract}
\begin{figure}[ht!]
\centering
\includegraphics[width=0.9\linewidth]{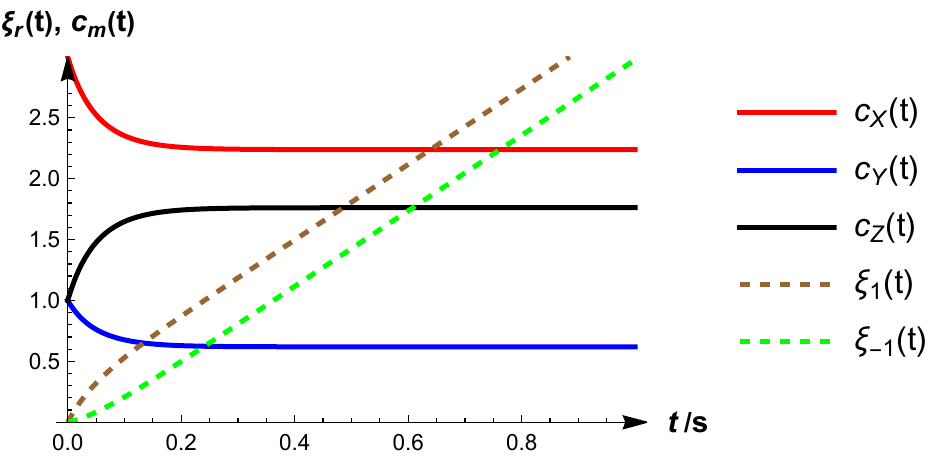}
\end{figure}

While the quantities tend to and are immediately very close to the stationary values, 
the reaction extents tend in a monotonously increasing way to infinity in the reaction: 
\ce{2X + Y <=>  2Z}.

\end{document}